\documentclass[aps,prb,twocolumn,reprint,superscriptaddress,floatfix]{revtex4-2}
\usepackage{graphicx}
\usepackage{dsfont}
\usepackage{amssymb}
\usepackage{hyperref}

\usepackage{xcolor}

\usepackage[acronym]{glossaries}

\newcommand{\ve}[1]{\boldsymbol{#1}}

\begin{document}
\title{Photoemission tomography of excitons in 2D systems: momentum-space signatures of correlated electron-hole wave functions}

\author{Siegfried Kaidisch}
\affiliation{Institute of Physics, NAWI Graz, University of Graz, 8010 Graz, Austria}

\author{Amir Kleiner}
\author{Sivan Refaely-Abramson}
\affiliation{Department of Molecular Chemistry and Materials Science, Weizmann Institute of Science, Rehovot, 7610001, Israel}

\author{Peter Puschnig}
\affiliation{Institute of Physics, NAWI Graz, University of Graz, 8010 Graz, Austria}

\author{Christian S. Kern}
\email[email: ]{christian.kern@uni-graz.at}
\affiliation{Institute of Physics, NAWI Graz, University of Graz, 8010 Graz, Austria}

\date{\today}

\begin{abstract}
The momentum-space signatures of excitons can be experimentally accessed through time-resolved (pump-probe) photoelectron spectroscopy. In this work, we develop a computational framework for exciton photoemission orbital tomography (exPOT) in periodic systems, enabling the simulation and interpretation of experimental observables within many-body perturbation theory. By connecting the $GW$+Bethe-Salpeter equation (BSE) approach to photoemission tomography, our formalism captures exciton photoemission in periodic systems, explicitly incorporating photoemission matrix element effects induced by the light-matter interaction via the probe pulse. The correlated nature of electrons and holes introduces distinct consequences for excitonic photoemission.
Using the prototypical two-dimensional material hexagonal boron nitride, we demonstrate these effects, including a dependence of the photoemission angular distribution on the pump pulse polarization.
Moreover, our framework extends to excitons with finite center-of-mass momentum, making it well-suited to studying momentum-dark excitons. This provides valuable insights into the microscopic nature of excitonic phenomena in quantum materials.
\end{abstract}

\maketitle

\section{\label{sec:intro}Introduction}
Excitons---bound electron-hole pairs---play a fundamental role in determining the optical and electronic response of semiconducting and insulating (two-dimensional) materials. From a theoretical perspective, describing excitonic effects requires going beyond single-particle approximations and typically involves time-dependent density functional theory (TDDFT) with suitable long-range xc-kernels, or many-body perturbation theory (MBPT) within the $GW$ plus Bethe-Salpeter equation (BSE) framework~\cite{Hybertsen1986,Rohlfing2000a,Onida2002}. Excitation energies and optical spectra computed in such a way can then be compared to absorption or luminescence spectroscopy. While theory naturally provides detailed information about the quantum mechanical nature of the underlying exciton wave function in real and momentum space, optical experiments provide only limited microscopic insights.

Here, time- and angle-resolved photoemission spectroscopy (trARPES) has emerged as a promising tool to dynamically probe excited states in energy and momentum space, thus offering enhanced insight into the microscopic nature of excitons~\cite{Madeo2020,Man2021,Dong2021,Wallauer2021,Wallauer2021a,Bennecke2024,Reutzel2024}. It is therefore highly desirable to have a theory that can connect trARPES signatures with first-principles calculations. The arguably most direct way of simulating trARPES is a numerical photodetection scheme, as implemented in real-time TDDFT~\cite{Dauth2016b,DeGiovannini2017}, which not only allows for pump-probe simulations in linear and nonlinear regimes but also captures final-state effects in the photoemission process~\cite{Dauth2016a,Kern2023}. Unfortunately, the only practically applicable approximation for the xc-kernel in such real-time TDDFT simulations is the adiabatic local density approximation, which fails to capture excitonic effects.

This is in contrast to Green's function methods, where the inclusion of Coulomb correlation between electron and hole, in conjunction with the screened exchange interaction, successfully describes bound electron-hole states. The non-equilibrium Greens function (NEGF) method is particularly powerful for simulating excited systems through explicit time propagation, capturing the effects of many-body interactions, such as electron-electron and electron-phonon scattering, via self-energy corrections~\cite{Stefanucci2025}. The photo\-emission signal is then derived from the lesser Green's function and the matrix element~\cite{Freericks2009}, with the former resulting from the non-equilibrium occupation due to the pump pulse and the latter accounting for the effects of the probe pulse. However, due to the complexity of the method, trARPES simulations within the NEGF framework have so far been restricted to model systems or Wannier-Hamiltonians~\cite{Perfetto2016,Schueler2021}, often neglecting the transition probability to the photoelectron state induced by the probe pulse~\cite{Rustagi2018,Sangalli2021}.

In one-step models for photoemission, these effects are encoded in the photoemission matrix element. Matrix element effects are not only crucial for accurately describing the photoemission process itself, but also encode rich information about quantum geometrical effects~\cite{Torma2023,Kang2025}, such as the Berry curvature~\cite{Cho2018,Schueler2020} or orbital angular momentum~\cite{Unzelmann2021,Schueler2020a}. Moreover, in organic or hybrid organic/inorganic systems, matrix element effects may easily become the dominant contribution to the observed angular dependence in the photoemission signal, forming the basis of the successful photoemission orbital tomography (POT) approach~\cite{Adawi1964,Feibelman1974,Puschnig2009a}.

For organic molecules, POT has recently been conceptually extended to exciton photoemission orbital tomography (exPOT)~\cite{Kern2023a}, which establishes a direct connection between the angular distributions observed in trARPES and the real-space and momentum-space structure of excitonic wave functions. However, this earlier work was limited to non-periodic systems, such as isolated molecules, thereby excluding its applicability to condensed-phase materials and low-dimensional crystals.

In the present work, we introduce exPOT for periodic systems, an ab initio scheme to simulate trARPES from first-principles. In contrast to explicit time-propagation methods, here we work in frequency space and with the exciton wave function from $GW$+BSE calculations in the linear response regime. The photoemission signal is then expressed in terms of the Fourier-transformed single-particle Bloch functions, weighted by the BSE eigenvectors and thus reflecting the entangled many-body character of electron-hole correlation. Our numerical implementation (see Appendix~\ref{sec:impl_details}) builds upon solutions of the BSE expressed in a plane wave basis set and is thus suited to treat crystalline solids and 2D semiconductors in the repeated-slab approach. We use the plane wave approximation for the final state in the photoemission process, including effects of the light polarization. 
We therefore expect our approach to be applicable for probe fields in the linear response regime and for probe energies in the 15--50~eV range.
Note that our generalized theory is also able to describe excitons with finite center-of-mass momentum, allowing for the treatment of optically dark excitons---a key feature in transition metal dichalcogenides and other layered materials~\cite{Ye2014a,Wu2015,Poellmann2015,Wallauer2021a}.  The exPOT approach in periodic systems thus enables access to a new class of excitonic observables in photoemission~\cite{Theilen2025}, bridging the gap between theoretical many-body wave functions and experimental momentum-resolved spectra.

This work is structured as follows. First we formulate the theoretical foundations in Sec.~\ref{sec:theory}, starting from Fermi's golden rule for the photoemission process, and including the many-body excitonic effects in terms of a Dyson orbital-like construction for Bloch functions. For energy-degenerate excitons, the resulting expression for photoemission from excitons also carries an additional dependence on the pump polarization, which is discussed thereafter. As an exemplification, we present results in Sec.~\ref{sec:results}, where we apply our method to monolayer hexagonal boron nitride (hBN), a prototypical 2D semiconductor with well-characterized excitonic properties. We further show that the resulting photoelectron angular distribution maps depend not only on the character of the exciton but also on the polarization of the pump pulse, and discuss the implications of our theory on the rich information contained in exciton photoemission.

\section{\label{sec:theory}Theory}
\subsection{Derivation of exPOT for periodic systems}
In this section, we derive the formalism on which we base our interpretation of photoelectron spectroscopy in terms of exciton wave functions and under the following conditions. First, we assume that the optical excitation (i.e. the pump pulse) is weak, such that we stay in the linear regime and well below the exciton Mott transition threshold. Moreover, the pump pulse is considered to be long enough to minimize energy broadening. Second, we assume that the probe pulse interacts with the system only after the pump pulse has ended. This ensures the preparation of a relaxed exciton~\cite{Sangalli2018,Marini2022} and avoids energy streaking effects in the trARPES spectrum induced by the pump field.  

We consider systems with a finite band gap and model the system's excited state using a correlated electron-hole wave function, expressed in terms of single-particle Bloch states as the basis set. To distinguish between electron and hole states, we write the valence state wave functions as $\phi_{v, \ve q}(\ve r)$, with $v$ denoting the band index and $\ve q$ the crystal momentum. Similarly, conduction state wave functions are written as $\xi_{c, \ve q}(\ve r)$. With these single particle basis sets and employing the Tamm-Dancoff approximation (TDA), we can write the $m$-th exciton wave function, $\Psi^{(m, \ve Q)}(\ve r_h, \ve r_e)$, in terms of electron- and hole coordinates, $\ve r_e$ and $\ve r_h$ respectively~\cite{Rohlfing2000a}:
\begin{align}
  \label{eq:ex_wave_bloch}
  \Psi^{(m, \ve Q)}(\ve r_h, \ve r_e)=
  \sum_{\ve q}^{\mathrm{BZ}} \sum_v^{N_{\mathrm{occ.}}} \sum_c^{\infty} X_{v, c, \ve q}^{(m, \ve Q)}
  \phi_{v, \ve q}^*(\ve r_h) \xi_{c, \ve q + \ve Q}(\ve r_e) 
\end{align}
In this expression, $\ve Q$ represents the center-of-mass momentum of the exciton, and $X_{v, c, \ve q}^{(m, \ve Q)}$ encode the exciton character. They are the elements of the transition density matrix, which we obtain as eigenstates from the BSE Hamiltonian, with the corresponding excitation energies denoted as $\Omega^{(m, \ve Q)}$. In the language of second quantization, the excitonic state may also be written as
\begin{align}
  \label{eq:initial_state}
  |\Psi_{m, \ve Q}^{N}\rangle = \sum_{\ve q}^{\mathrm{BZ}} \sum_{v, c} X_{v, c, \ve q}^{(m, \ve Q)} 
  \hat{a}_{c, \ve q+\ve Q}^{\dagger} \hat{a}_{v, \ve q} |\Psi_{0}^N\rangle.
\end{align}
Here, $\hat{a}_{v, \ve q}$ annihilates an electron with crystal momentum $\ve q$ in the valence band $v$ (or creates a hole) in the ground state $|\Psi_{0}^N\rangle$, and $\hat{a}_{c, \ve q+\ve Q}^{\dagger}$ creates an electron in the conduction band $c$ with momentum $\ve q+\ve Q$. 

In analogy to ground state POT~\cite{Puschnig2009a}, and under the conditions stated above, we treat the photoemission due to the probe pulse in first-order perturbation theory. As outlined in Ref.~\cite{Dauth2014}, we apply Fermi's golden rule for the transition rate from an $N$-electron initial state $|\Psi_\mathrm{i}^N\rangle$ to a final state $|\Psi_\mathrm{f}^N\rangle$: 
\begin{align}
\label{eq:goldenrule}
W_{\mathrm{i} \rightarrow \mathrm{f}} = 2 \pi \left| \left\langle \Psi_\mathrm{f}^N \right| \ve{A} \ve{\hat{P}} \left| \Psi_\mathrm{i}^N \right\rangle \right|^2
 \delta\left(\omega + E_\mathrm{i} - E_\mathrm{f} \right),
\end{align}
with the Dirac delta function enforcing energy conservation between the initial and final state and the energy $\omega$ carried by the photon field $\ve A$. Here and in the following, we use the velocity gauge within the dipole approximation and use atomic units unless denoted otherwise.

The final state $|\Psi_\mathrm{f}^{N}\rangle$ describes the system after the photoemission event. Here, we make use of the sudden approximation~\cite{Damascelli2004} and thus neglect correlations between the photoelectron described by the one-electron state $|\gamma_{\ve k}\rangle$ and the remaining $(N-1)$-electron state $|\Psi_\mathrm{f}^{N-1}\rangle$. 
Therefore, $|\Psi_\mathrm{f}^N\rangle$ can be written as an anti-symmetrized product:
\begin{align}
  |\Psi_\mathrm{f}^N \rangle = 
  \mathcal{\hat{A}} (|\Psi_\mathrm{f}^{N-1}\rangle \otimes |\gamma_{\ve k}\rangle)
  \equiv 
  |\mathcal{A}\Psi_\mathrm{f}^{N-1} \gamma_{\ve k}\rangle,
\end{align}
where we additionally assume the $(N-1)$-electron final state to be of the form
\begin{align}
  \label{eq:final_N-1}
  |\Psi_{j,{\ve{\tilde{q}}}}^{N-1}\rangle \equiv 
  \hat{a}_{j, \ve{\tilde{q}}}|\Psi_{0}^N\rangle,
\end{align}
with $\hat{a}_{j, \ve{\tilde{q}}}$ annihilating an electron (creating a photo-hole) in band $j$, and with crystal momentum $\ve{\tilde{q}}$, from the $N$-electron ground state $|\Psi_{0}^N\rangle$. Via this approximation, we neglect possible relaxations in the final $(N-1)$-electron system.

In the following, we take the $m$-th excited state with center-of-mass momentum $\ve Q$ as defined in Eq.~\ref{eq:initial_state} as initial state, and assume that in the final state, the emitted electron has momentum $\ve k$, while the photo-hole resides in $|\phi_{j,\ve{\tilde{q}}}\rangle$.
Then, the matrix element $\left\langle \Psi_\mathrm{f}^N \right| \ve{A} \ve{\hat{P}} \left| \Psi_\mathrm{i}^N \right\rangle$ of Eq.~\ref{eq:goldenrule} takes the form
\begin{align}
  \label{eq:matrix_element1}
  M^\mathrm{(m,\ve{Q})}_{j,\ve{\tilde{q}},\ve k}(\ve A) \equiv
  \langle \mathcal{A} \Psi_{j,\ve{\tilde{q}}}^{N-1} \gamma_{\ve k} | \ve{A} \ve{\hat{P}}| \Psi_\mathrm{m,\ve{Q}}^N \rangle.
\end{align}
Note that with this matrix element, we can calculate the transition rate from $|\Psi_\mathrm{m,\ve{Q}}^N\rangle$ to \emph{one} specific final state, $|\mathcal{A} \Psi_{j,\ve{\tilde{q}}}^{N-1} \gamma_{\ve k}\rangle$ where an electron has been removed from the $j$-th band (and carrying crystal momentum $\ve{\tilde{q}}$). To account for all possible photoemission final states, we will have to sum up the transition rates of all final hole quantum numbers at a later stage.

To proceed, we reduce the $N$-electron matrix element of Eq.~\ref{eq:matrix_element1} to an effective one-electron matrix element by introducing the Dyson orbital as the overlap between the initial $N$-electron and the final $(N-1)$-electron system~\cite{Dauth2014,Truhlar2019,Krylov2020}. In contrast to existing literature for isolated systems, we here perform this step for Bloch functions that carry an additional crystal momentum quantum number. We then get for the absolute of the matrix element
\begin{align}
  \label{eq:matrix_element2}
  \left| M^\mathrm{(m,\ve{Q})}_{j,\ve{\tilde{q}},\ve k}(\ve A) \right| =  
  \left| \int_V \mathrm d^3 r \gamma_{\ve k}^*(\ve r) \ve{A} \ve{\nabla} D_{j,\ve{\tilde{q}}}^{(m, \ve Q)}(\ve r) \right|,
\end{align}
where we have defined the effective single-particle Dyson wave function with Bloch momentum $\ve{\tilde{q}}$ as
\begin{align}
  \label{eq:dyson1}
D_{j,\ve{\tilde{q}}}^{(m, \ve Q)}(\ve r) = 
&\sqrt{N} \int_V \mathrm d^3 r_2 ... \mathrm d^3 r_{N}
\langle \Psi_{j,\ve{\tilde{q}}}^{N-1} | 
\ve r_2,...,\ve r_{N} \rangle& \nonumber \\
&\cdot \langle \ve r, \ve r_2,...,\ve r_{N} | 
\Psi_{\mathrm m, \ve Q}^N\rangle.
\end{align}
Here the volume of integration, $V$, is the volume of the crystal, as defined by the unit cell and the sampling of the Brillouin-zone in the periodic directions (using sufficient vacuum in the non-periodic direction of the slab).

Next we evaluate the formal definition of the Dyson wave function, starting from an alternative form, which employs the basis of valence- and conduction bands and is derived in Appendix~\ref{sec:dyon_expansion}:
\begin{align} 
  \label{eq:dyson_expanded}
  D_{j,\ve{\tilde{q}}}^{(m, \ve Q)}(\ve r) = 
  \sum_{v'}\sum_{\ve q'}^{\mathrm{BZ}}
  \langle \Psi_{j,\ve{\tilde{q}}}^{N-1} | 
  \hat{a}_{v', \ve q'}
  | \Psi_{m, \ve Q}^N    \rangle 
  \phi_{v',\ve q'}(\ve r) &\nonumber \\
  +
  \sum_{c'}\sum_{\ve q'}^{\mathrm{BZ}}
  \langle \Psi_{j,\ve{\tilde{q}}}^{N-1}  | 
  \hat{a}_{c', \ve q'}
  | \Psi_{m, \ve Q}^N \rangle 
  \xi_{c',\ve q'}(\ve r) &
\end{align}
Inserting Eqs.~\ref{eq:initial_state} and \ref{eq:final_N-1}, we get
\begin{align} 
  \label{eq:dyson2}
  &D_{j,\ve{\tilde{q}}}^{(m, \ve Q)}(\ve r) = \\
  &=\sum_{v,v',c}\sum_{\ve q, \ve q'}^{\mathrm{BZ}}
  X_{v, c, \ve q}^{(m, \ve Q)}
  \langle\Psi_{0}^N|
  \hat{a}_{j, \ve{\tilde{q}}}^{\dagger}\hat{a}_{v', \ve q'} \hat{a}_{c, \ve q+\ve Q}^{\dagger}  \hat{a}_{v, \ve q}
  | \Psi_{0}^N\rangle 
  \phi_{v',\ve q'}(\ve r)
  \nonumber \\
  &+\sum_{v,c,c'}\sum_{\ve q, \ve q'}^{\mathrm{BZ}}
  X_{v, c, \ve q}^{(m, \ve Q)}
  \langle\Psi_{0}^N|
  \hat{a}_{j, \ve{\tilde{q}}}^{\dagger}\hat{a}_{c', \ve q'} \hat{a}_{c, \ve q+\ve Q}^{\dagger} \hat{a}_{v, \ve q}
  |\Psi_{0}^N\rangle 
  \xi_{c',\ve q'}(\ve r).\nonumber
\end{align}
For simplicity, we assume the ground state, $|\Psi_{0}^N\rangle$, to be a single Slater-determinant constructed from the Kohn-Sham valence orbitals $\phi_{v,\ve q}$. This allows us to work with the $GW$+BSE method carried out on top of density functional theory (DFT) calculations, but we stress that our method is also suitable for more general, correlated initial states. The first term in Eq.~\ref{eq:dyson2} containing the sum over $v'$ then vanishes due to orthogonality of the Bloch states. By the same argument, we can set $c'=c$, $v=j$, $\ve q=\ve{\tilde{q}}$ and $\ve q'=\ve q+\ve Q$ in the second term and thus find
\begin{align}
  \label{eq:dyson3}
  D_{v,\ve{q}}^{(m, \ve Q)}(\ve r) = \sum_{c}
  X_{v, c, \ve{q}}^{(m, \ve Q)}
  \xi_{c,\ve{q}+\ve Q}(\ve r).
\end{align}
Inserting Eq.~\ref{eq:dyson3} into Eq.~\ref{eq:matrix_element2}, we arrive at
\begin{align}
  \label{eq:matrix_element3}
  \left|M^\mathrm{(m,\ve{Q})}_{v,\ve q, \ve k}(\ve A)\right|
  = 
  \left|\int_V \mathrm d^3 r \gamma_{\ve k}^*(\ve r) \ve{A} \ve{\nabla} 
  \sum_{c}
  X_{v, c, \ve{q}}^{(m, \ve Q)}
  \xi_{c,\ve{q}+\ve Q}(\ve r)\right|.
\end{align}

To further evaluate the matrix element, we follow the commonly applied strategy in POT and approximate the photoelectron wave function by a plane wave final state (PWFS): $\gamma_{\ve k}(\ve r)=e^{\mathrm i \ve k \ve r}/\sqrt{V}$. Integrating by parts, and dropping the surface term, then allows us to write the matrix element as a sum over Fourier-transformed Bloch functions:
\begin{align}
  \label{eq:matrix_element4}
  \left|M^\mathrm{(m,\ve{Q})}_{v,\ve q, \ve k}(\ve A) \right|
  =
  \left| \frac{\ve A \ve k}{\sqrt{V}}
  \sum_{c}
  X_{v, c, \ve{q}}^{(m, \ve Q)}
  \mathcal F [\xi_{c,\ve{q}+\ve Q}](\ve k)\right|
\end{align}

Before putting everything together, we also take the energy conservation defined in Eq.~\ref{eq:goldenrule} into account. To this end, we identify the energy of the initial state as the sum of the ground state energy, $E_0^N$, and the excitation energy $\Omega^{(m, \ve Q)}$. 
The energy of the final state consists of the photoelectron kinetic energy, $E_\mathrm{kin}=\ve k^2/2$ and the energy of the $(N-1)$-electron state, $E_{v,\ve{q}}^{N-1}$. The latter can be expressed as the ground-state energy plus the ionization energy of the $v$-th valence band at $\ve{q}$: $\varepsilon_{v, \ve{q}}$. This leads to
\begin{align}
  \label{eq:energy_conservation}
  \delta\left(\omega + E_\mathrm{i} - E_\mathrm{f} \right)
  = \delta\left(\omega + \Omega^{(m, \ve Q)} -\varepsilon_{v, \ve{q}} - E_\mathrm{kin}\right).
\end{align}

Combining this expression for the energy conservation with the matrix element of Eq.~\ref{eq:matrix_element4}, we can now rewrite the transition rate $W$ in Eq.~\ref{eq:goldenrule}. This rate gives the probability for the $m$-th exciton, with excitation energy $\Omega^{(m, \ve Q)}$ and center-of-mass momentum $\ve Q$, to end up in a state with photoelectron momentum $\ve k$, if the photohole resides in valence orbital $\phi_{v,\ve q}$:
\begin{align}
W_{(m,\ve Q) \rightarrow (v, \ve{q},\ve{k})} 
= &\frac{2 \pi}{V}
|\ve A \ve k|^2
\left|\sum_{c} X_{v, c, \ve{q}}^{(m, \ve Q)}\mathcal F [\xi_{c,\ve{q}+\ve Q}](\ve k)\right|^2 
\nonumber \\
&\cdot \delta\left(\omega + \Omega^{(m, \ve Q)} -\varepsilon_{v, \ve{q}} - E_\mathrm{kin}\right)
\end{align}
Summing up the transition rates for all final-state quantum-numbers $v$ and $\ve{q}$, we get an expression which is proportional to the overall photoelectron intensity, 
\begin{align}
  \label{eq:finalintensity}
  I_{m,\ve Q}(\ve k) \propto &|\ve A \ve k|^2 \sum_v \sum_{\ve{q}}^{\mathrm{BZ}}\left|
  \sum_{c} X_{v, c, \ve{q}}^{(m, \ve Q)}\mathcal F [\xi_{c,\ve{q}+\ve Q}](\ve k)\right|^2 \nonumber \\
  &\cdot \delta\left(\omega + \Omega^{(m, \ve Q)} -\varepsilon_{v, \ve{q}} - E_\mathrm{kin}\right),
\end{align}
where $I_{m,\ve Q}(\ve k)$ denotes the number of photoelectrons of momentum $\ve k$, that are emitted from the $m$-th excited state carrying center-of-mass momentum $\ve Q$.

In summary, photoemission from a static (or, in practice, long-lived, quasi-static) exciton state is given by a sum over crystal momentum and hole state contributions, for which the photoelectron kinetic energy is determined by the energy conservation between the probe-, pump-, and hole energy. At each allowed kinetic energy, the photoelectron angular distribution is given by the absolute squared of a coherent sum over electron contributions, weighted by the respective electron-hole transition coefficients, which are obtained from the solutions of the BSE. In contrast to other formulations of exciton photoemission that only consider the spectral function \cite{Rustagi2018,Sangalli2021}, our formalism includes photoelectron matrix element effects due to the probe pulse and at the level of the PWFS approximation.

\subsection{Pump-induced effects on exciton photoemission} \label{subsec:pump_pol}
In the preceding derivation, we assumed the system to be in an \emph{eigenstate} of the BSE Hamiltonian, $|m, \ve Q\rangle$, whose real-space representation is given in Eq.~\ref{eq:ex_wave_bloch}. The resulting expression for the photoelectron intensity (Eq.~\ref{eq:finalintensity}) applies to such an eigenstate and does not consider how the system was initially prepared by the pump pulse. In general however, the pump pulse typically generates a \emph{superposition} state---a linear combination of the calculated eigenstates---which necessitates an adaptation of our formalism, as detailed below. In preparation for Sec.~\ref{sec:results}, we restrict the discussion to optical excitations ($\ve Q \rightarrow 0$), allowing us to denote the eigenstates of the BSE Hamiltonian as $|m, \ve Q=0\rangle=:|m\rangle$.
The superposition state $|S_l\rangle$, created by the pump laser, then becomes
\begin{align}
  |S_l\rangle = \sum_{m} c_m^l |m\rangle.
\end{align}

As mentioned above, the presented formalism Eq.~\ref{eq:finalintensity} is only suitable for photoemission from an eigenstate of the BSE Hamiltonian. 
Assuming that the superposition state $|S_l\rangle$, created by the pump field, is a combination of energy-degenerate eigenstates (obtained from solving the BSE), it is itself an eigenstate of the BSE Hamiltonian, thus allowing us to still apply the formalism.
Therefore, we define a small enough energy difference $\eta > 0$, which spans the degenerate subspace around excitation energy $\Omega_l$, and get the expansion coefficients $c_m^l$, dependent on the polarization $\ve E$ of the pump pulse, via the optical transition matrix elements:
\begin{align}
  c_m^l \propto \langle 0|\ve E \cdot \hat{\ve v}|m\rangle\; \delta\left(|\Omega^{(m)} - \Omega_l| < \eta \right).
\end{align}
Here $|0\rangle$ denotes the ground state and $\hat{\ve v} = \mathrm i [\hat{H}, \hat{\ve r}]$ the single-particle velocity operator. 

As a consequence, for photoemission intensity from the superposition (eigen)state $|S_l\rangle$, also the transition density matrix elements are linearly combined and thus Eq.~\ref{eq:finalintensity} must be modified accordingly ($\ve Q=0$):
\begin{align}
  \label{eq:finalintensity_mixed}
  I_{S_l}(\ve k) \propto &|\ve A \ve k|^2 \sum_v \sum_{\ve{q}}^{\mathrm{BZ}}\left|
  \sum_{c} \sum_m c^l_m X_{v, c, \ve{q}}^{(m)}\mathcal F [\xi_{c,\ve{q}}](\ve k)\right|^2 \nonumber \\
  &\cdot \delta\left(\omega + \Omega_l -\varepsilon_{v, \ve{q}} - E_\mathrm{kin}\right)
\end{align}
Note, that the new transition density matrix $\sum_m c^l_m X_{v, c, \ve{q}}^{(m)}$ can readily be identified as the BSE eigenvector of the superposition state. The procedure presented in this section can thus also be understood as a simple basis change within the subspace of degenerate eigenstates, or, in other words, as a change from the basis obtained through solution of the BSE to a basis determined by the pump polarization.

In 2D systems excitons typically occur as degenerate pairs, and only their superposition should be considered as the physical exciton, which then, for instance, also respects the symmetries of the system~\cite{Galvani2016,UriaAlvarez2024}. In contrast to those eigenstates, the pump pulse in general breaks the system's symmetry and may create superimposed exciton states that depend on the pump polarization. As we will show in Sec.~\ref{subsec:pump_pol_results}, such pump polarization effects have a considerable imprint on exciton photoemission. 

\section{\label{sec:results}Results}
\subsection{Calculation of hBN quasiparticle and optical properties}
\begin{figure}
  \includegraphics[width=8.5cm]{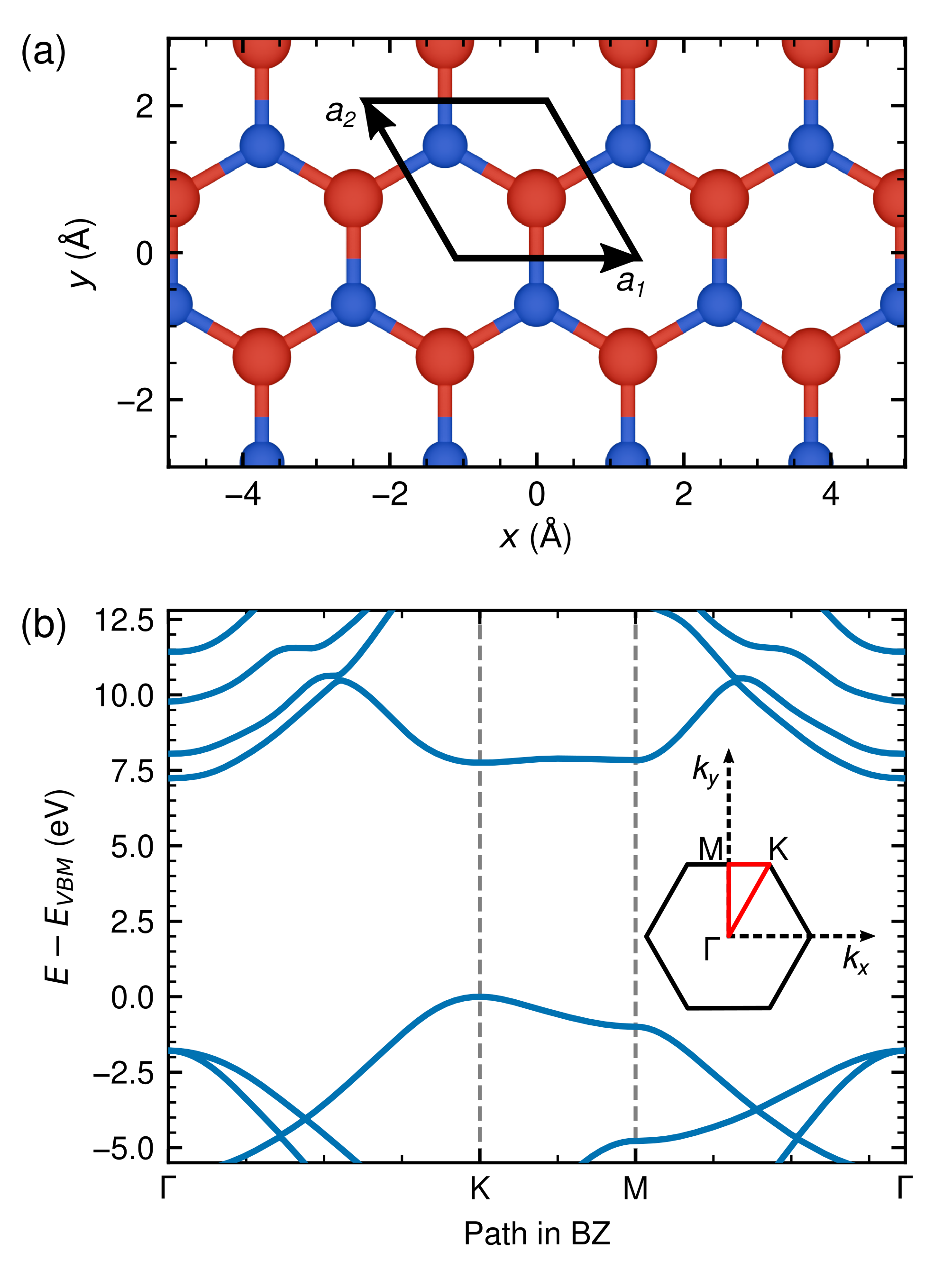}
  \caption{
  (a) Atomic structure of hBN. Boron and nitrogen atoms are depicted in red and blue, respectively. The vectors $\mathbf{a_1}$ and $\mathbf{a_2}$ span the primitive unit cell, which contains one atom of each kind. (b) $G_0W_0$ band structure of hBN. Quasiparticle energies are shown along the $\Gamma - K - M - \Gamma$ high symmetry path. The calculated band structure exhibits an indirect band-gap, with the valence-band maximum and conduction-band minimum being located at $K$ (and $K'$) and $\Gamma$, respectively.}
  \label{fig:structure}
\end{figure}
In the following, we exemplify our formalism by the simulation of excitonic photoemission in monolayer hexagonal boron nitride (hBN) as a prototypical 2D system. As a starting point, we obtain wave functions and energy levels from DFT calculations with \textsc{Quantum Espresso}~\cite{Giannozzi2009, Giannozzi2017, Giannozzi2020}, using the PBE-GGA exchange-correlation functional~\cite{Perdew1996} and a lattice constant of 2.507~\AA, see Fig.~\ref{fig:structure}~(a). To compensate for the systematic underestimation of the band gap within semi-local DFT, we apply the $G_0W_0$ method to correct the energy levels. Utilizing the \textsc{BerkeleyGW} code~\cite{Hybertsen1986, Rohlfing2000a, Deslippe2012}, we obtain a direct (indirect) band gap of 7.6~eV (7.4~eV), which is comparable to literature values from similar methods, see e.g. Ref.~\cite{Kirchhoff2022} for a comparison. We show the $G_0W_0$ band structure in Fig.~\ref{fig:structure}~(b) and detail the computational methods in Appendix~\ref{sec:comp_details}.
\begin{figure}
  \includegraphics[width=8.5cm]{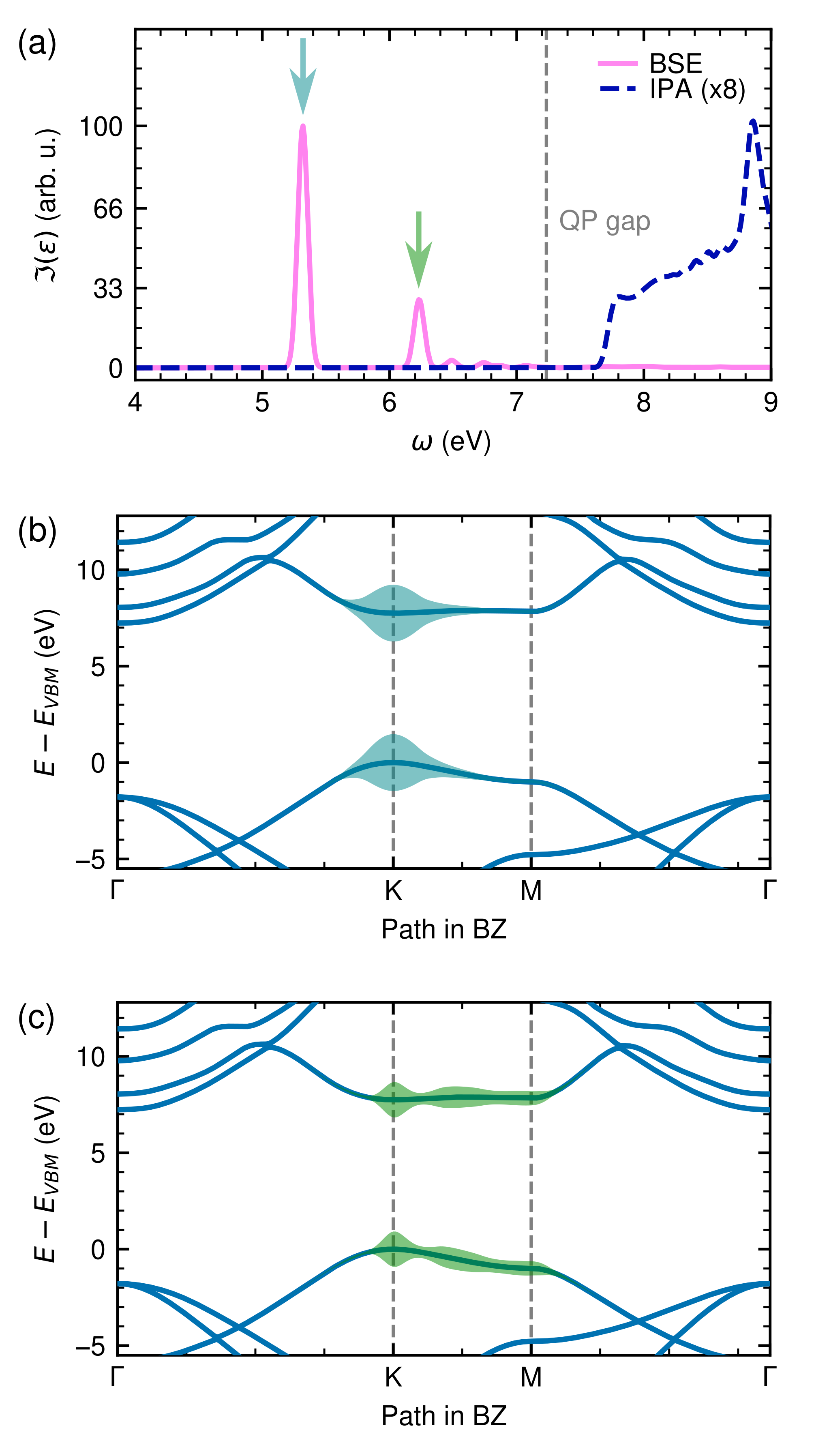}
  \caption{Optical excitation of monolayer hBN. (a) Independent-particle spectrum, based on a $G_0W_0$ calculation in dark blue and the BSE spectrum in pink. In the latter, $S_1$ and $S_2$ energies are marked by arrows.
  Panels (b) and (c) show the $G_0W_0$ band structure with overlays indicating the BSE eigenvector composition of the $S_1$ (blue) and $S_2$ (green) excitations of the system, respectively.}
  \label{fig:bse}
\end{figure}

Subsequently, we use the DFT wave functions and $G_0W_0$ energy levels, and solve the BSE within the TDA and for $\ve Q=0$. This yields the excitation energies $\Omega_m$ and the BSE eigenvectors $|m,\ve Q=0\rangle \equiv |m\rangle$, which we use to simulate optical spectra via Eq.~\ref{eq:finalintensity}. As shown in Fig.~\ref{fig:bse} (a), the BSE spectrum features a range of excitonic peaks below the band gap. These are often interpreted as hydrogen-like solutions of the Wannier equation~\cite{wannier1937structure}, though they have also been found to deviate from that model~\cite{Galvani2016}. For comparison, we include the optical spectrum computed within the independent-particle approximation, shown as a dashed blue curve in Fig.~\ref{fig:bse} (a).
In the following, we focus on the two lowest excitons, located at 5.3~eV ($S_1$) and 6.2~eV ($S_2$). 

Both, $S_1$ and $S_2$ are composed of two degenerate eigenstates of the BSE Hamiltonian. This results in an additional dependence on the pump polarization, as discussed in Sec.~\ref{subsec:pump_pol}, and which will become relevant at a later stage in Sec.~\ref{subsec:pump_pol_results}. For $S_1$ at 5.3~eV, the two degenerate valence-conduction transitions occur at $K$ and $K'$, respectively, with their momentum-dependent transition weights indicated in Fig.~\ref{fig:bse}~(b). The situation is similar for $S_2$ at 6.2~eV, albeit with transition weights being more spread out along the Brillouin-zone boundary, as can be seen in Fig.~\ref{fig:bse}~(c).

\subsection{Exciton photoemission in hBN}
\begin{figure}
  \includegraphics[width=8.5cm]{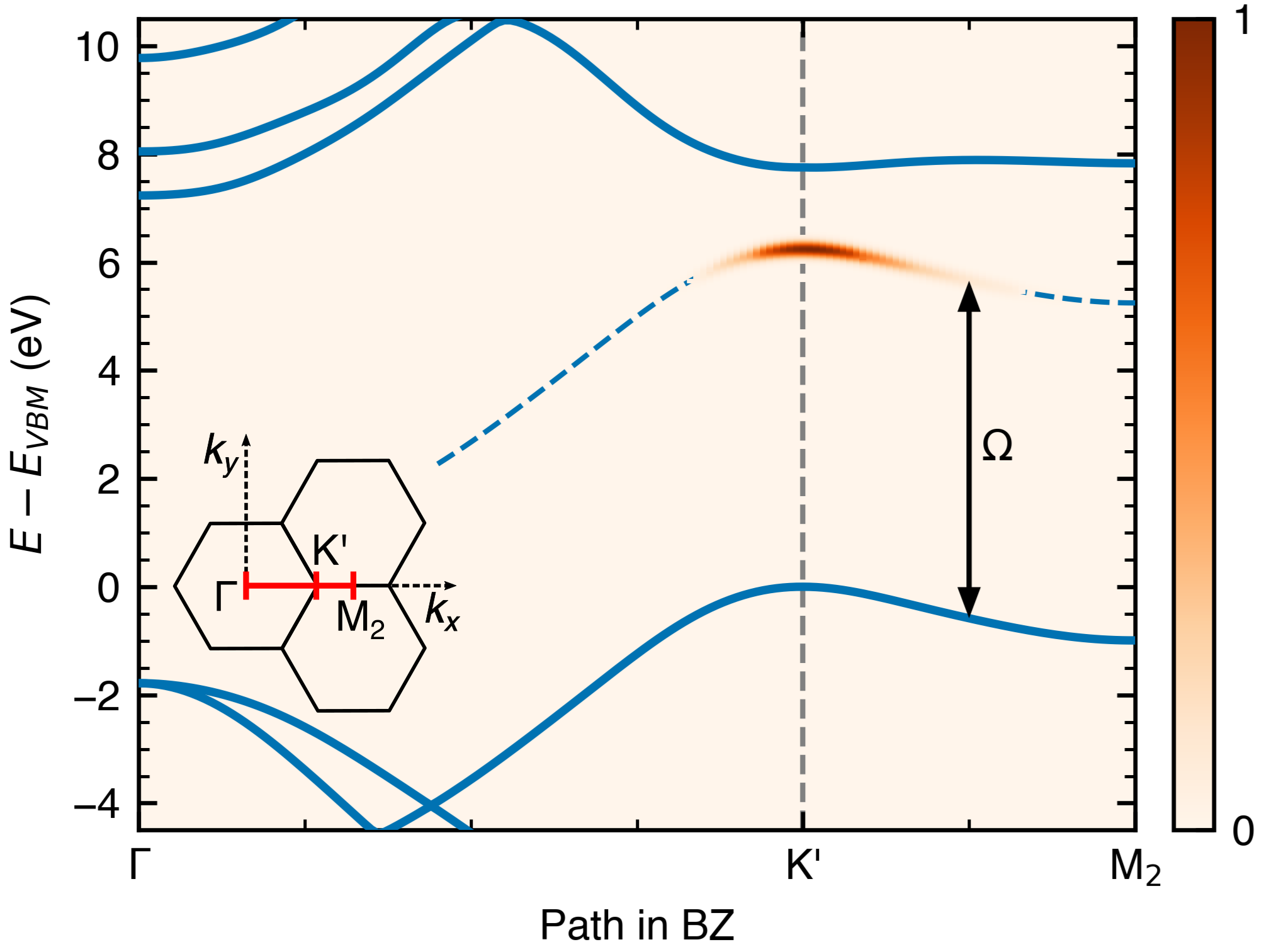}
  \caption{ARPES bandmap for the $S_1$ exciton in $k_x$ direction (cut of the BZ along $\Gamma-K'-M_2$, see inset with three neighboring BZs). The $G_0W_0$ bandstructure is overlaid in full blue lines, with a duplicate of the valence band, shifted by $\Omega$, as a blue dashed line.}
  \label{fig:darkcorridor}
\end{figure}

With the quasiparticle and optical properties at hand, we now illustrate consequences of our theory for exciton photoemission from monolayer hBN. In the following, we use a 26.5~eV probe energy with a Gaussian energy broadening of 0.1~eV. For a better insight into the pump-induced effects on the momentum maps, we ignore the trivial effect of probe polarization, stemming from the $|\ve A \ve k|^2$ prefactor in Eq.~\ref{eq:finalintensity}.
Note, that our formulation only considers excitonic photoemission, thereby neglecting photoemission from the initially populated electronic structure and associated effects like its depopulation due to exciton creation.
This decoupling is a direct consequence of energy conservation in excitonic photoemission Eq.~\ref{eq:finalintensity}. The exciton energy $\Omega^{(m,\mathbf Q)}$ is transferred to the photoelectron, yielding higher kinetic energies than those obtained from photoemission out of the initially populated electronic states (see Fig.~\ref{fig:darkcorridor}).

In Fig.~\ref{fig:darkcorridor} we show the simulated bandmap for photoemission from the $S_1$ exciton together with the $G_0W_0$ bandstructure (full blue lines) along the $\Gamma-K'-M_2$ path (corresponding to the $k_x$ direction, see inset). According to Eq.~\ref{eq:finalintensity}, the energy dispersion of exciton photoemission is determined by the hole component of the exciton. Consequently, the exciton's trARPES signature appears at the photoelectron kinetic energy corresponding to the valence band energy plus the pump energy $\Omega$ (blue dashed line). This finding, which is well-established in the literature~\cite{Weinelt2004,Rustagi2018}, enables the estimation of the exciton's spatial extent from photoemission data under the assumption of an $s$-like envelope function~\cite{Madeo2020,Man2021,Karni2022}.

\begin{figure}[ht]
  \includegraphics[width=8.5cm]{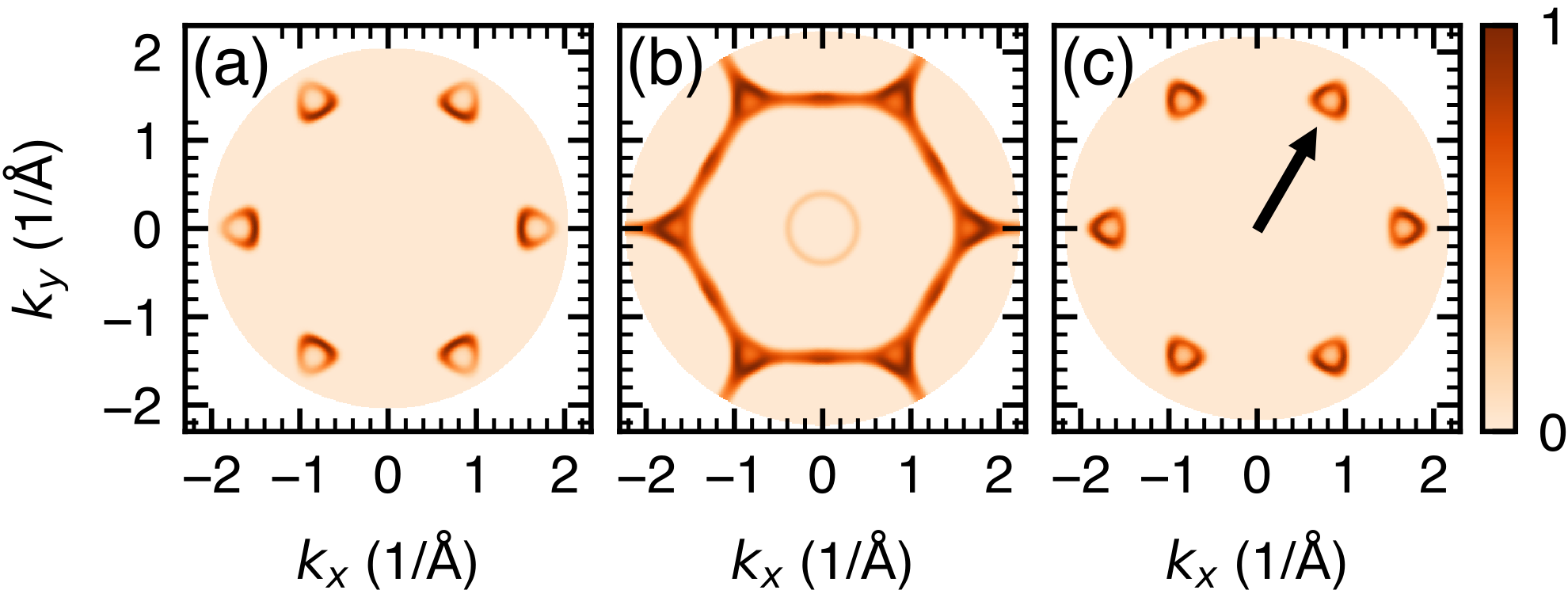}
  \caption{Momentum maps for three different cases: (a) valence band at $E_b=E_{\mathrm{VBM}}-0.2$~eV ($E_{\mathrm{kin}}=18.6$~eV), (b) conduction band at $E_b=E_{\mathrm{CBM}}+0.6$~eV ($E_{\mathrm{kin}}=26.6$~eV), and (c) $S_1$ exciton at $E_b=E_{\mathrm{VBM}}+\Omega-0.2$~eV ($E_{\mathrm{kin}}=23.9$~eV).
  }
  \label{fig:vbm_cbm_ex}
\end{figure}
Next, it is expedient to compare momentum maps of exciton photoemission with those for photoemission from the ground state, which we obtain by the standard way of simulating POT~\cite{Lueftner2017}. In Fig.~\ref{fig:vbm_cbm_ex}~(a), we present such a map for a kinetic energy slightly below the valence band maximum (VBM) ($E-E_{\mathrm{VBM}}=-0.2$~eV), which is comprised of six triangular features, each around the $K$/$K'$ points. This pattern is analogous to the case of graphene, where the triangular features are also broken in the outgoing direction of the Brillouin-zone, leading to croissant- or horseshoe-like shapes. This ``dark corridor'' is due to matrix element effects in the photoemission process and has already been described in great detail for the case of graphene~\cite{Gierz2011,Krasovskii2021,Kern2023}. When assuming a static, charge-transfer induced population of the conduction band, the latter would also be accessible to photoemission from the ground state, which we simulate in panel~(b) of Fig.~\ref{fig:vbm_cbm_ex}, using a kinetic energy of 26.6~eV, which is above the conduction band minimum (CBM) ($E-E_{\mathrm{CBM}}=+0.6$~eV). In contrast to the valence band case depicted in panel~(a), the features around the Dirac point are now elongated in the directions going to the $M$/$M'$ points, which is reasonable when looking at the bandstructure (see e.g. Fig.~\ref{fig:structure}). Remarkably, the dark corridor from the matrix element effect is now pointing to the \emph{inside} of the BZ. Returning now to the simulation of exciton photoemission, we show a momentum map of the $S_1$ exciton in Fig.~\ref{fig:vbm_cbm_ex}~(c), using a kinetic energy of 23.9~eV and a pump polarization as indicated by a black arrow. As already discussed in the preceding paragraph, the exciton inherits the energy dispersion of the \emph{hole} (here: the highest valence band), but, in light of Eq.~\ref{eq:finalintensity}, follows the momentum distribution given by the Fourier-transformed \emph{electron} part of the exciton wave function (here: the lowest conduction band). As a direct imprint of the latter, the dark corridor is also pointing to the inside of the BZ, which we can see clearly in Fig.~\ref{fig:vbm_cbm_ex}~(c). Having established the unoccupied part of the electronic structure as another direct marker of exciton photoemission, we will now examine the different contributions to the momentum maps in more detail.

\begin{figure*}
  \includegraphics[width=15cm]{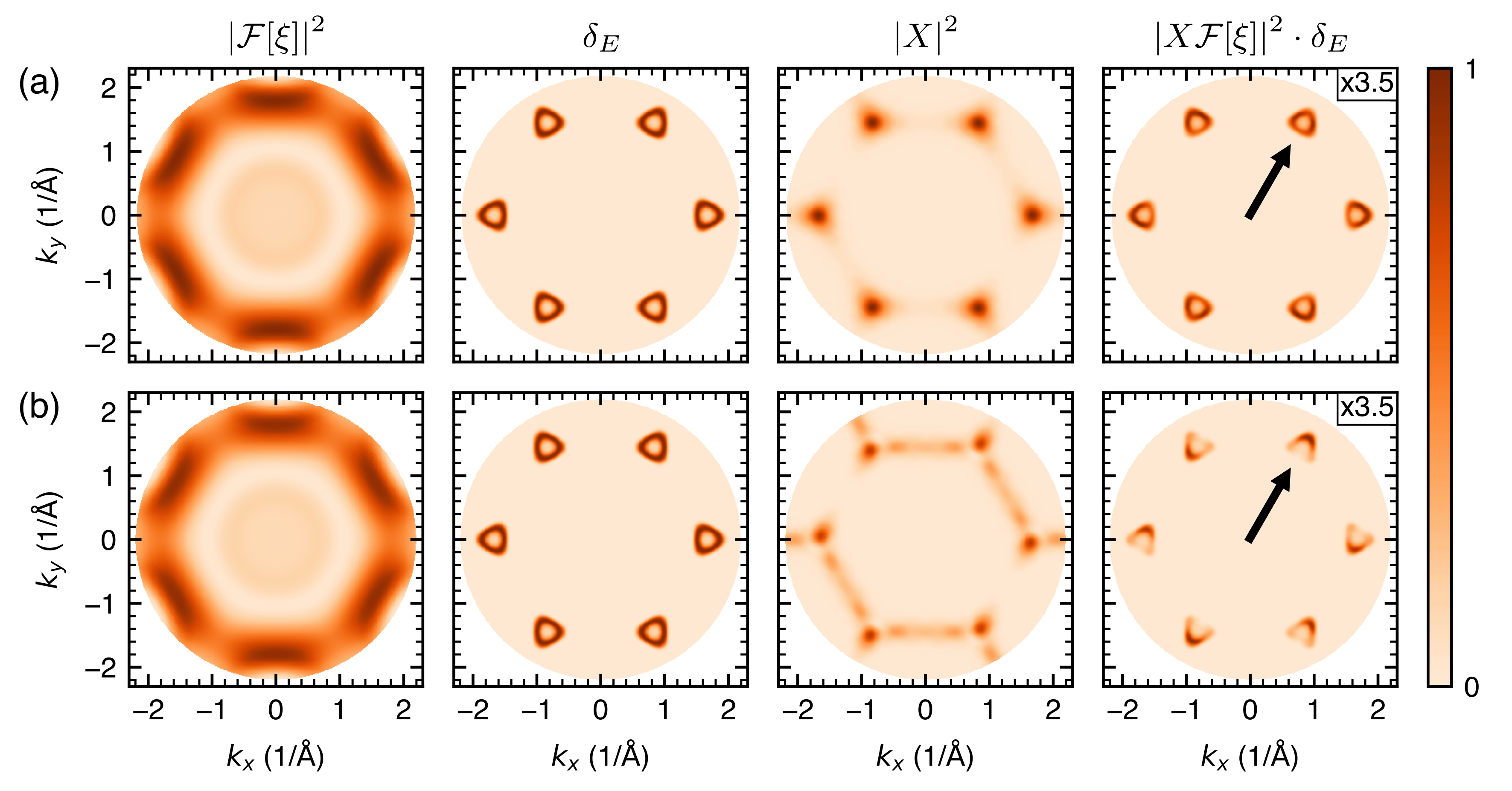}
  \caption{Decomposition of exciton ARPES maps for $S_1$ in row~(a) and $S_2$ in row~(b), with black arrows indicating the polarization of the pump laser. First column: Fourier-transformed conduction-band wave functions. Second column: energy-conservation delta-function. Third column: BSE eigenvectors. Fourth column: resulting exciton ARPES maps as products of first to third columns. Each map is individually normalized to one. For $S_1$ in row (a), the horseshoes point away from $\Gamma$, with minor modulations along their contours. For $S_2$ in row (b), four of the six horseshoes are rotated, when compared to the first exciton. This is a direct result of the differences in the momentum-space distribution of their BSE eigenvectors.}
  \label{fig:composition}
\end{figure*}
Using Eq.~\ref{eq:finalintensity}, we dissect the photoemission maps for both excitons, $S_1$ and $S_2$, in order to understand the influence of each contributing factor.
Since---in this case---both excitons only involve transitions from the highest valence to the lowest conduction band, the sums over $v$ and $c$ in Eq.~\ref{eq:finalintensity} collapse. Consequently, we are left with a single product of $|\mathcal{F}[\xi]|^2$, $\delta_E$ and $|X|^2$, involving only wave functions from the lowest conduction band, energies of the highest valence band and BSE transition coefficients from the highest valence band to the lowest conduction band. Fig.~\ref{fig:composition} shows these three factors, as well as the full photoemission maps, for the two excitons in row~(a) ($S_1$) and row~(b) ($S_2$), respectively. The direction of the pump polarization is indicated in the rightmost column. For both excitons $S_1$ and $S_2$, the triangular patterns are generated by the $\delta_E$ factor, with $|\mathcal{F}[\xi]|^2$ only leading to slight modifications. The maps in the first two columns, $|\mathcal{F}[\xi]|^2$ and $\delta_E$, are very similar between the two excitons $S_1$ and $S_2$, with minor differences arising due to the unequal photoelectron kinetic energies (23.9~eV for $S_1$ and 24.8~eV for $S_2$). Since all other areas already have vanishing intensity, the additional multiplication with the BSE eigenvector can at most lead to intensity differences within the area of the horseshoes. As can be seen from the third column, the BSE eigenvectors of the two excitons differ, leading to the differences in their respective photoemission maps (rightmost column of Fig.~\ref{fig:composition}). In the following, those differences are inspected more closely. 

For $S_1$, the BSE eigenvector (third column in Fig.~\ref{fig:composition}~(a)) mostly exhibits a six-fold symmetry. It is centered around the $K$ and $K'$ points and varies weakly along the contours of the horseshoes, thus leading only to small intensity variations (fourth column in Fig.~\ref{fig:composition}~(a)) along them. 
Contrary, in the case of $S_2$ in row~(b), the BSE eigenvector exhibits a nodal line perpendicular to the pump polarization --- which has broken the six-fold symmetry --- while being non-zero along the remaining four sides of the hexagonal BZ.
Since the pump polarization was chosen along a high-symmetry direction (indicated by black arrows in the rightmost column), it acts as a symmetry axis for the BSE eigenvector and, consequently, for the resulting photoemission map.
In contrast to the case of $S_1$, the regions around $K$ and $K'$ show much more details for $S_2$, particularly featuring small nodal lines, which, upon multiplication of the BSE eigenvector with the horseshoe pattern produced by $|\mathcal{F}[\xi]|^2$ and $\delta_E$, suppress photoemission intensity. 
Moreover, the local maxima are shifted off $K$ and $K'$. As can be seen in the last column of Fig.~\ref{fig:composition}, the combination of all three factors effectively leads to a redistribution of spectral intensity within the horseshoes. By selectively accessing the different factors in Eq.~\ref{eq:finalintensity}, we are thus able to deconvolute different contributions to the momentum maps. In the general case of multiple transitions within an exciton, however, the sums over different valence and conduction bands remain and, as a result, the above analysis would become less straightforward, especially due to the coherent summation of the electron part. Since in the present case of the hBN example we found the most asymmetry to arise due to the BSE eigenvectors in conjunction with the pump polarization, we will discuss this effect in more detail in the following.

\subsection{Pump polarization effects in 2D hBN}\label{subsec:pump_pol_results}
\begin{figure*}[htb]
  \includegraphics[width=15cm]{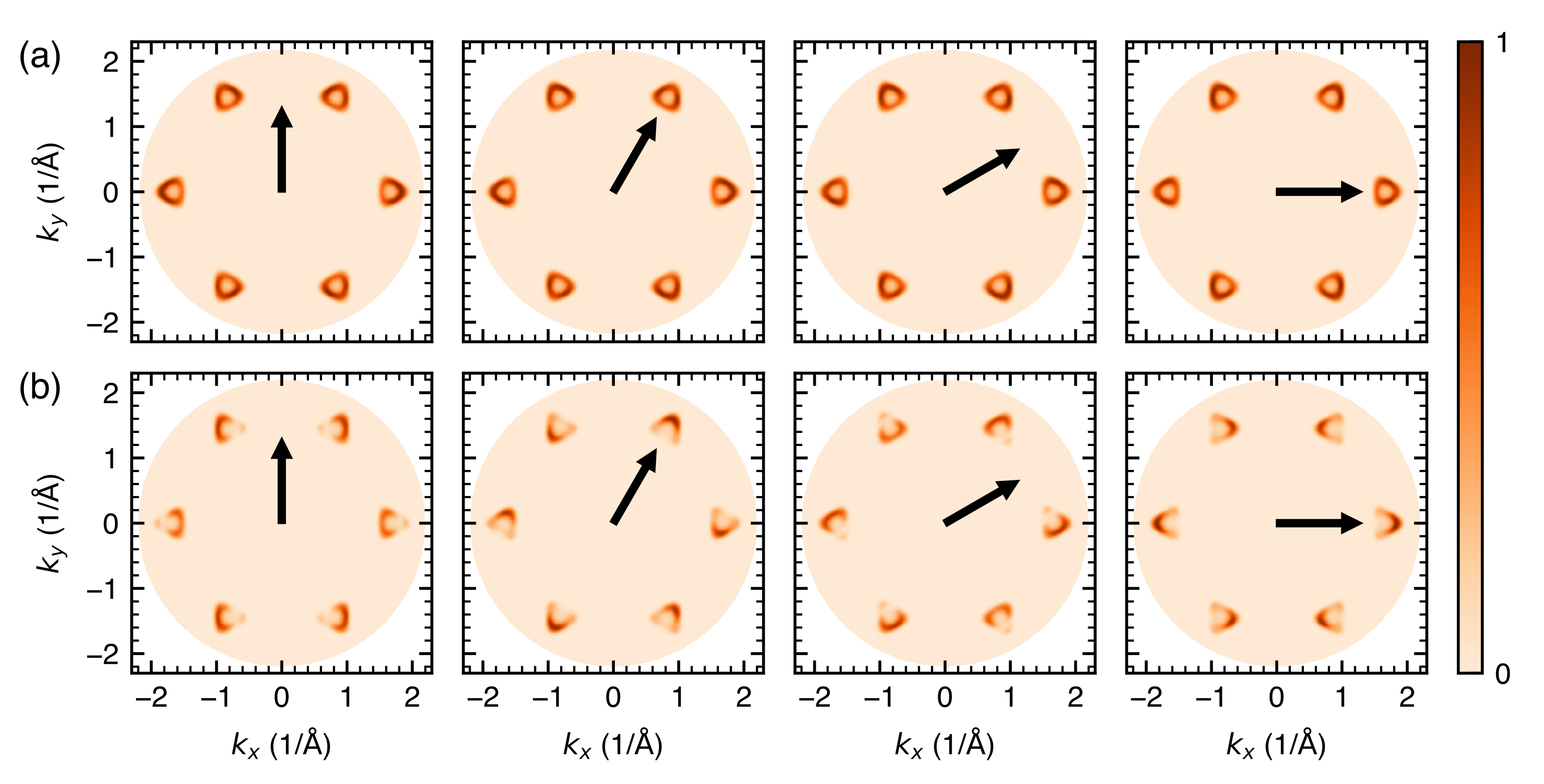}
  \caption{Influence of pump laser's polarization on exciton photoemission.
  The four columns shows ARPES photoemission maps for different polarizations of the pump laser (black arrows).
  (a) shows photoemission from the lowest-energy exciton while (b) shows the second exciton.
  In the case of the first exciton, only small differences in the intensity distribution along the horseshoes is visible, while for the second exciton, there is a strong dependence on the pump polarization, with different polarizations leading to different rotations of the horseshoes.
  In the case of both excitons, the photoemission maps are mirror symmetric across the (high-symmetry) polarization directions. 
  }
  \label{fig:polarization}
\end{figure*}
As outlined in Sec.~\ref{subsec:pump_pol}, the summation of energy-degenerate BSE eigenvectors, in order to form a superimposed exciton, involves coefficients that depend on the polarization of the pump pulse.
The influence of pump polarization is illustrated in detail in Fig.~\ref{fig:polarization}. Panels (a) and (b) in Fig.~\ref{fig:polarization} depict the photoemission from the $S_1$ exciton (5.3~eV) at a kinetic energy of 23.9~eV and the $S_2$ exciton (6.2~eV) at 24.8~eV, respectively. The black arrows indicate the polarization direction of the pump pulse. Photoemission from the first exciton is only weakly dependent on the pump pulse's polarization and retains the horseshoe intensity pattern. Varying the polarization reveals minor variations in the intensity distributions along the horseshoes. For the second exciton, however, there is a strong dependence of the photoemission intensity distribution on the pump pulse, where different polarization directions lead to different redistributions of spectral weight within the horseshoes. Although the probe pulse polarization may introduce an overall attenuation in an trARPES experiment, pump-induced effects on exciton photoemission are therefore expected to be observable, provided that sufficient energy and momentum resolution are achieved.

\section{Conclusions}
We have developed a framework for exciton photoemission orbital tomography (exPOT) in periodic systems, thereby enabling the computation and interpretation of photoemission angular distributions from excitons in crystalline two-dimensional materials. The exPOT formalism is derived from first principles, using many-body perturbation theory and the Dyson orbital concept, and connects the photoemission intensity to results from $GW$+Bethe-Salpeter calculations. As a manifestation of the electron-hole correlation, for each hole configuration, exciton photoemission is expressed as a coherent sum over Fourier-transformed conduction-band wave functions, weighted by the Bethe–Salpeter eigenvector components.

We have implemented the formalism in a computational workflow compatible with standard plane wave basis set $GW$+BSE, using DFT wave functions, $GW$ energies, and BSE eigenvectors. As a proof of concept, we have applied our formalism to the prototypical case of monolayer hexagonal boron nitride. The resulting exPOT maps exhibit distinct patterns for different excitonic states. These patterns are characterized by an energy dispersion that follows the hole states, while the electron part of the exciton wave function is reflected in the photoemission matrix element. Both contributions are further modified by the BSE eigenvectors. The latter introduces an additional dependence on the pump polarization, resulting in the suppression of spectral weight in the momentum maps as a direct consequence of the excitonic character.

The exPOT formalism for periodic systems provides a powerful tool for gaining microscopic insight into the internal structure of excitons and offers a predictive framework for designing and interpreting next-generation trARPES experiments on quantum materials. Our formalism also includes excitons with finite center-of-mass momentum, making it particularly suited for elucidating the nature of momentum-dark excitons by bridging the gap between ab-initio many-body theory and experimentally measurable momentum-resolved spectra.

\clearpage
\appendix

\section{Expansion of Dyson orbitals in Bloch states}
\label{sec:dyon_expansion}
In this section, we derive the expansion Eq.~\ref{eq:dyson_expanded} of the Dyson orbitals, as given in Eq.~\ref{eq:dyson1}, in terms of Bloch functions.

Since this expansion holds in general, we start with the Dyson orbital formed by the overlap of an arbitrary $(N-1)$- and $N$-electron wave function:
\begin{align}
D(\ve r) = 
\sqrt{N} \int_V \mathrm d^3 r_2 ... \mathrm d^3 r_{N}
\Psi^{N-1}
\left(\ve r_2,...,\ve r_{N} \right)^* &\nonumber \\
\Psi^N
\left(\ve r, \ve r_2,...,\ve r_{N}\right).&
\end{align}

Next, since the Dyson-orbitals are one-electron wave functions, they can be expanded in any set of one-electron basis functions. For a set of (Born-von Kármán-periodic, BvK-periodic, just like $D$) orthonormal basis functions $\{ \psi_j(\cdot) \}$ we can thus write
\begin{align} 
  \label{eq:basis_expansion1}
  D(\ve r) = 
  &\sum_{j} c_j \psi_j(\ve r),
\end{align}
with linear coefficients
\begin{align}
  c_j = \int_V d^3r \psi_j(\ve r)^* D(\ve r),
\end{align}
where $V$ is the crystal volume, see text below Eq.~\ref{eq:dyson1}.
Inserting the Dyson orbital and renaming $\ve r$ to $\ve r_1$, we get
\begin{align}
  c_j = &\sqrt{N}\int_V \mathrm d^3 r_1 \mathrm d^3 r_2 ... \mathrm d^3 r_{N}
  \psi_j(\ve r_1)^*  \nonumber \\
  & \Psi^{N-1}
  \left(\ve r_2,...,\ve r_{N} \right)^* 
  \Psi^N
  \left(\ve r_1, \ve r_2,...,\ve r_{N}\right).
\end{align}
Next, we rename $\ve r_1$ to $\ve r_2$ and vice versa to get 
\begin{align}
  c_j =& \sqrt{N}\int_V \mathrm d^3 r_1 \mathrm d^3 r_2 ... \mathrm d^3 r_{N}
  \psi_j(\ve r_2)^* \nonumber \\
  & \Psi^{N-1}
  \left(\ve r_1,\ve r_3,...,\ve r_{N} \right)^*  
  \Psi^N
  \left(\ve r_2, \ve r_1,...,\ve r_{N}\right) \nonumber \\
  =& \sqrt{N}\int_V \mathrm d^3 r_1 \mathrm d^3 r_2 ... \mathrm d^3 r_{N}
  \psi_j(\ve r_2)^*   \nonumber \\
  & \Psi^{N-1}
  \left(\ve r_1,\ve r_3,...,\ve r_{N} \right)^*
  \left(-\Psi^N
  \left(\ve r_1, \ve r_2,...,\ve r_{N}\right)\right),
\end{align}
where we used the Pauli exclusion principle in the last step.
Using the same argument, we find for any $i$ in $\{1,...,N\}$:
\begin{align}
  c_j =& \sqrt{N}\int_V \mathrm d^3 r_1 ... \mathrm d^3 r_{N}
  (-1)^{1+i}
  \psi_j(\ve r_i)^*   \nonumber \\
  & \Psi^{N-1}
  \left(\ve r_1,..., \ve r_{i-1}, \ve r_{i+1},...,\ve r_{N} \right)^*
  \Psi^N
  \left(\ve r_1, \ve r_2,...,\ve r_{N}\right).
\end{align}
We can thus write $c_j$ as a sum over $i$ as:
\begin{align}
  c_j &= \frac{1}{N} \sum_{i=1}^{N} 
  \sqrt{N}\int_V \mathrm d^3 r_1 ... \mathrm d^3 r_{N}
  (-1)^{1+i}
  \psi_j(\ve r_i)^*   \nonumber \\
  & \quad \Psi^{N-1}
  \left(\ve r_1,..., \ve r_{i-1}, \ve r_{i+1},...,\ve r_{N} \right)^*
  \Psi^N
  \left(\ve r_1,...,\ve r_{N}\right) \nonumber \\
  &= \int_V \mathrm d^3 r_1 ... \mathrm d^3 r_{N}
  \Psi^N\left(\ve r_1,...,\ve r_{N}\right) \nonumber \\
  & \cdot \left(
  \sum_{i=1}^{N} 
  \frac{(-1)^{1+i}}{\sqrt{N}}
  \psi_j(\ve r_i)
  \Psi^{N-1}\left(\ve r_1,..., \ve r_{i-1}, \ve r_{i+1},...,\ve r_{N} \right)
  \right)^*
\end{align}
Now we can identify the expression in parentheses with the wave function of the anti-symmetrized product state $|{\mathcal A} \psi_j \Psi^{N-1} \rangle$ (see, e.g., Ref.~\cite{Dauth2014}). 
Using the creation operator $\hat{a}_{j}^{\dagger}$ corresponding to $|\psi_j\rangle$ and a bra-ket notation, we can thus write
\begin{align}
  c_j =& 
  \int_V \mathrm d^3 r_1 ... \mathrm d^3 r_{N}
  \langle \ve r_1,...,\ve r_N | \Psi^{N} \rangle \nonumber \\
  &\left(
  \langle \ve r_1,...,\ve r_N | \hat{a}_{j}^{\dagger} | \Psi^{N-1} \rangle
  \right)^* \nonumber \\
  =& \int_V \mathrm d^3 r_1 ... \mathrm d^3 r_{N} 
  \langle \Psi^{N-1} | \hat{a}_{j} | \ve r_1,...,\ve r_N \rangle
  \langle \ve r_1,...,\ve r_N | \Psi^{N} \rangle.
\end{align}

Finally, we want to arrive at a basis-independent representation of $c_j$, which allows for efficient handling of the ladder-operator matrix-elements.
To do so, we make use of the fact, that for every wave function $\Psi(\cdot)$, that satisfies BvK periodic boundary conditions for a large crystal, there is a non-periodic wave function $\tilde{\Psi}(\cdot)$, that is equal to $\Psi(\cdot)$ inside the crystal, but vanishes quickly outside of the crystal.
Then, we get 
\begin{align}
  c_j 
  &= \int_V \mathrm d^3 r_1 ... \mathrm d^3 r_{N} 
  \langle \Psi^{N-1} | \hat{a}_{j} | \ve r_1,...,\ve r_N \rangle
  \langle \ve r_1,...,\ve r_N | \Psi^{N} \rangle \nonumber \\
  &\approx \int_V \mathrm d^3 r_1 ... \mathrm d^3 r_{N} 
  \langle \tilde{\Psi}^{N-1} | \hat{\tilde{a}}_{j} | \ve r_1,...,\ve r_N \rangle
  \langle \ve r_1,...,\ve r_N | \tilde{\Psi}^{N} \rangle, 
\end{align}
where $\hat{\tilde{a}}_{j}$ is the destruction operator for $|\tilde{\psi}_j\rangle$, which is the non-periodic equivalent of basis vector $|\psi_j\rangle$.
Assuming the non-periodic functions vanish quickly outside of $V$, we then get
\begin{align}
  c_j 
  &\approx \int_{\mathds{R}^3} \mathrm d^3 r_1 ... \mathrm d^3 r_{N} 
  \langle \tilde{\Psi}^{N-1} | \hat{\tilde{a}}_{j} | \ve r_1,...,\ve r_N \rangle
  \langle \ve r_1,...,\ve r_N | \tilde{\Psi}^{N} \rangle \nonumber \\
  &= \langle \tilde{\Psi}^{N-1} | \hat{\tilde{a}}_{j} | \tilde{\Psi}^{N} \rangle,
\end{align}
where we used 
\begin{align}
\hat{\mathds{1}} = \int_{\mathds{R}^3} d^3 r |\ve{r}\rangle\langle \ve{r}|.
\end{align}
We can thus perform operations in a basis-independent manner using the non-periodic wave functions and then transform back to the BvK wave functions. 
Symbolically, we can thus simply write 
\begin{align}
  c_j 
  &= \langle \Psi^{N-1} | \hat{a}_{j} | \Psi^{N} \rangle
\end{align}
as a shorthand for the above-shown steps.

Inserting this into Eq.~\ref{eq:basis_expansion1}, we get our final result for the expansion of the Dyson orbital $D$ in an arbitrary, orthonormal basis set $\{\psi_j(\cdot)\}$:
\begin{align} 
  D(\ve r) = 
  &\sum_{j} \langle \Psi^{N-1} | \hat{a}_{j} | \Psi^{N} \rangle \psi_j(\ve r).
\end{align}
To get to Eq.~\ref{eq:dyson_expanded}, we then simply use the Bloch functions as basis functions and split the sum over bands into a sum over valence bands and a sum over conduction bands.

\section{Implementation Details}\label{sec:impl_details}

This section provides Eq.~\ref{eq:finalintensity} in an alternative form that allows for efficient implementation when using a plane-wave basis set. First note, that in Eq.~\ref{eq:finalintensity} in the Brillouin-zone (BZ) summation over the Bloch momentum $\ve q$, only a single $\ve q$ vector contributes. 
This can be seen, without loss of generality, by expanding $\xi_{c,\ve{q}+\ve Q}(\ve r)$ in a plane-wave basis (using Born-von Kármán, BvK, periodic boundary conditions):
\begin{align}
  \xi_{c,\ve{q}+\ve Q}(\ve r) &= \frac{1}{V} \sum_{\ve G} c_{c,\ve{q}+\ve Q}(\ve G) e^{\mathrm{i}(\ve q + \ve Q + \ve G)\ve r}, \nonumber \\
  c_{c,\ve{q}+\ve Q}(\ve G) &= \int_V \mathrm{d}^3 r \xi_{c,\ve{q}+\ve Q}(\ve r) e^{-\mathrm{i}(\ve q + \ve Q + \ve G)\ve r}.
\end{align}
The Fourier-transform
\begin{align}
\mathcal F [\xi_{c,\ve{q}+\ve Q}](\ve k) = \int_V \mathrm{d}^3 r \xi_{c,\ve{q}+\ve Q}(\ve r) e^{-\mathrm{i}\ve k \ve r}
\end{align}
then vanishes unless $\ve q + \ve G + \ve Q = \ve k $ for at least one of the reciprocal lattice vectors $\ve G$.
Now, since $\ve q$ is restricted to the BZ and $\ve G$ is a reciprocal lattice vector, there is at most a single combination of $\ve q$ and $\ve G$ that satisfies this condition.
Thus, the BZ summation is reduced to a single contributing $\ve q$ vector and 
\begin{align}
  \mathcal F [\xi_{c,\ve{q}+\ve Q}](\ve k) = \sum_{\ve G} c_{c,\ve{q}+\ve Q}(\ve G) \delta(\ve q + \ve G+ \ve Q - \ve k ). 
\end{align}

The photoemission intensity may then be written as
\begin{align}
  \label{eq:finalintensity2}
  &I_{m,\ve Q}(\ve k) \propto |\ve A \ve k|^2 \sum_v \sum_{\ve{q}}^{\mathrm{BZ}}\sum_{\ve{G}}
  \delta(\ve q + \ve G+ \ve Q - \ve k )  \nonumber \\
  &\cdot \left|\sum_{c} X_{v, c, \ve{q}}^{(m, \ve Q)} c_{c,\ve{q}+\ve Q}(\ve G)\right|^2 
  \delta\left(\omega + \Omega^{(m, \ve Q)} -\varepsilon_{v, \ve{q} - E_\mathrm{kin}}\right).
\end{align}
This can be used for an efficient implementation of the presented formalism: 
The grid of $\ve G$ vectors, determined by the used unit cell and the grid of $\ve q$ vectors, determined by the used BZ sampling, create a grid of $\ve k$ vectors, $\ve k = \ve q + \ve G + \ve Q$, for which we get:
\begin{align}
  \label{eq:finalintensity3}
  I_{m,\ve Q}(\ve k) \propto &|\ve A \ve k|^2 \sum_v 
  \left|\sum_{c} X_{v, c, \ve{q}}^{(m, \ve Q)} c_{c,\ve{q}+\ve Q}(\ve G)\right|^2 \nonumber \\
  &\cdot \delta\left(\omega + \Omega^{(m, \ve Q)} -\varepsilon_{v, \ve{q}} - E_\mathrm{kin} \right), \nonumber \\
  & \mathrm{with}~\ve q + \ve G + \ve Q= \ve k.
\end{align}
For any photoelectron momentum $\ve k$ on this grid, we can thus directly use the plane-wave coefficients $c_{c,\ve{q}+\ve Q}(\ve G)$, BSE eigenvector components $X_{v, c, \ve{q}}^{(m, \ve Q)}$ and $GW$ energies $\varepsilon_{v, \ve{q}}$ to obtain the photoemission intensity $I_{m,\ve Q}(\ve k)$, without the need for a BZ summation over $\ve q$.
Given a sufficiently fine BZ sampling, the photoemission intensity for any $\ve k$ not on this grid (but within range of the energy cutoff used for wave functions in the DFT calculation) can then be obtained using efficient interpolation techniques.

\section{Computational Details}\label{sec:comp_details}
For the example of monolayer hBN, we calculate Kohn-Sham DFT wave functions and energies in \textsc{Quantum Espresso}~\cite{Giannozzi2009, Giannozzi2017, Giannozzi2020} using the Perdew-Burke-Ernzerhof exchange-correlation functional, ONCV pseudopotentials and a cutoff energy of 50~Ry and 200~Ry for the wave functions and the representation of the density, respectively. The lattice constant was set to 2.507~\AA (bulk experimental value: 2.506~\AA~\cite{Bosak2006}) and we use a vacuum layer of 20~\AA, as well as a truncation of the Coulomb interaction in $z$ direction~\cite{Sohier2017}, in order to avoid spurious interaction between repeated slabs from the periodic boundary condition.
Based on these ground-state properties, we perform a $G_0W_0$ calculation using \textsc{BerkeleyGW}~\cite{Hybertsen1986, Rohlfing2000a, Deslippe2012}. In particular, we use the \textit{NNS} subsampling technique, which, for systems with reduced dimensionality, accelerates convergence with respect to the Brillouin-zone sampling~\cite{da_Jornada2017}. For the calculation of the static inverse dielectric function we use a 12x12x1 sampling of the Brillouin-zone, a cutoff energy of 30~Ry for reciprocal lattice vectors and sum over 200 bands.
Based on this, we calculate the $G_0W_0$ energies for 4 valence and 44 conduction bands, using the same k-space mesh and reciprocal lattice cutoff as for the inverse dielectric function.

Finally, we construct the BSE Hamiltonian using 4 valence and 9 conduction bands on a 12x12x1 k-space mesh, interpolate it to a 96x96x1 k-space mesh (4 valence, 4 conduction bands) and diagonalize it.

\bibliography{references}

\begin{thebibliography}{57}%
\makeatletter
\providecommand \@ifxundefined [1]{%
 \@ifx{#1\undefined}
}%
\providecommand \@ifnum [1]{%
 \ifnum #1\expandafter \@firstoftwo
 \else \expandafter \@secondoftwo
 \fi
}%
\providecommand \@ifx [1]{%
 \ifx #1\expandafter \@firstoftwo
 \else \expandafter \@secondoftwo
 \fi
}%
\providecommand \natexlab [1]{#1}%
\providecommand \enquote  [1]{``#1''}%
\providecommand \bibnamefont  [1]{#1}%
\providecommand \bibfnamefont [1]{#1}%
\providecommand \citenamefont [1]{#1}%
\providecommand \href@noop [0]{\@secondoftwo}%
\providecommand \href [0]{\begingroup \@sanitize@url \@href}%
\providecommand \@href[1]{\@@startlink{#1}\@@href}%
\providecommand \@@href[1]{\endgroup#1\@@endlink}%
\providecommand \@sanitize@url [0]{\catcode `\\12\catcode `\$12\catcode
  `\&12\catcode `\#12\catcode `\^12\catcode `\_12\catcode `\%12\relax}%
\providecommand \@@startlink[1]{}%
\providecommand \@@endlink[0]{}%
\providecommand \url  [0]{\begingroup\@sanitize@url \@url }%
\providecommand \@url [1]{\endgroup\@href {#1}{\urlprefix }}%
\providecommand \urlprefix  [0]{URL }%
\providecommand \Eprint [0]{\href }%
\providecommand \doibase [0]{https://doi.org/}%
\providecommand \selectlanguage [0]{\@gobble}%
\providecommand \bibinfo  [0]{\@secondoftwo}%
\providecommand \bibfield  [0]{\@secondoftwo}%
\providecommand \translation [1]{[#1]}%
\providecommand \BibitemOpen [0]{}%
\providecommand \bibitemStop [0]{}%
\providecommand \bibitemNoStop [0]{.\EOS\space}%
\providecommand \EOS [0]{\spacefactor3000\relax}%
\providecommand \BibitemShut  [1]{\csname bibitem#1\endcsname}%
\let\auto@bib@innerbib\@empty
\bibitem [{\citenamefont {Hybertsen}\ and\ \citenamefont
  {Louie}(1986)}]{Hybertsen1986}%
  \BibitemOpen
  \bibfield  {author} {\bibinfo {author} {\bibfnamefont {M.~S.}\ \bibnamefont
  {Hybertsen}}\ and\ \bibinfo {author} {\bibfnamefont {S.~G.}\ \bibnamefont
  {Louie}},\ }\href {https://doi.org/10.1103/PhysRevB.34.5390} {\bibfield
  {journal} {\bibinfo  {journal} {Phys. Rev. B}\ }\textbf {\bibinfo {volume}
  {34}},\ \bibinfo {pages} {5390} (\bibinfo {year} {1986})}\BibitemShut
  {NoStop}%
\bibitem [{\citenamefont {Rohlfing}\ and\ \citenamefont
  {Louie}(2000)}]{Rohlfing2000a}%
  \BibitemOpen
  \bibfield  {author} {\bibinfo {author} {\bibfnamefont {M.}~\bibnamefont
  {Rohlfing}}\ and\ \bibinfo {author} {\bibfnamefont {S.~G.}\ \bibnamefont
  {Louie}},\ }\href {https://doi.org/10.1103/PhysRevB.62.4927} {\bibfield
  {journal} {\bibinfo  {journal} {Phys. Rev. B}\ }\textbf {\bibinfo {volume}
  {62}},\ \bibinfo {pages} {4927} (\bibinfo {year} {2000})}\BibitemShut
  {NoStop}%
\bibitem [{\citenamefont {Onida}\ \emph {et~al.}(2002)\citenamefont {Onida},
  \citenamefont {Reining},\ and\ \citenamefont {Rubio}}]{Onida2002}%
  \BibitemOpen
  \bibfield  {author} {\bibinfo {author} {\bibfnamefont {G.}~\bibnamefont
  {Onida}}, \bibinfo {author} {\bibfnamefont {L.}~\bibnamefont {Reining}},\
  and\ \bibinfo {author} {\bibfnamefont {A.}~\bibnamefont {Rubio}},\ }\href
  {https://doi.org/10.1103/RevModPhys.74.601} {\bibfield  {journal} {\bibinfo
  {journal} {Rev. Mod. Phys.}\ }\textbf {\bibinfo {volume} {74}},\ \bibinfo
  {pages} {601} (\bibinfo {year} {2002})}\BibitemShut {NoStop}%
\bibitem [{\citenamefont {Mad\'eo}\ \emph {et~al.}(2020)\citenamefont
  {Mad\'eo}, \citenamefont {Man}, \citenamefont {Sahoo}, \citenamefont
  {Campbell}, \citenamefont {Pareek}, \citenamefont {Wong}, \citenamefont
  {Al-Mahboob}, \citenamefont {Chan}, \citenamefont {Karmakar}, \citenamefont
  {Mariserla}, \citenamefont {Li}, \citenamefont {Heinz}, \citenamefont {Cao},\
  and\ \citenamefont {Dani}}]{Madeo2020}%
  \BibitemOpen
  \bibfield  {author} {\bibinfo {author} {\bibfnamefont {J.}~\bibnamefont
  {Mad\'eo}}, \bibinfo {author} {\bibfnamefont {M.~K.~L.}\ \bibnamefont {Man}},
  \bibinfo {author} {\bibfnamefont {C.}~\bibnamefont {Sahoo}}, \bibinfo
  {author} {\bibfnamefont {M.}~\bibnamefont {Campbell}}, \bibinfo {author}
  {\bibfnamefont {V.}~\bibnamefont {Pareek}}, \bibinfo {author} {\bibfnamefont
  {E.~L.}\ \bibnamefont {Wong}}, \bibinfo {author} {\bibfnamefont
  {A.}~\bibnamefont {Al-Mahboob}}, \bibinfo {author} {\bibfnamefont {N.~S.}\
  \bibnamefont {Chan}}, \bibinfo {author} {\bibfnamefont {A.}~\bibnamefont
  {Karmakar}}, \bibinfo {author} {\bibfnamefont {B.~M.~K.}\ \bibnamefont
  {Mariserla}}, \bibinfo {author} {\bibfnamefont {X.}~\bibnamefont {Li}},
  \bibinfo {author} {\bibfnamefont {T.~F.}\ \bibnamefont {Heinz}}, \bibinfo
  {author} {\bibfnamefont {T.}~\bibnamefont {Cao}},\ and\ \bibinfo {author}
  {\bibfnamefont {K.~M.}\ \bibnamefont {Dani}},\ }\href
  {https://doi.org/10.1126/science.aba1029} {\bibfield  {journal} {\bibinfo
  {journal} {Science}\ }\textbf {\bibinfo {volume} {370}},\ \bibinfo {pages}
  {1199} (\bibinfo {year} {2020})}\BibitemShut {NoStop}%
\bibitem [{\citenamefont {Man}\ \emph {et~al.}(2021)\citenamefont {Man},
  \citenamefont {Madéo}, \citenamefont {Sahoo}, \citenamefont {Xie},
  \citenamefont {Campbell}, \citenamefont {Pareek}, \citenamefont {Karmakar},
  \citenamefont {Wong}, \citenamefont {Al-Mahboob}, \citenamefont {Chan},
  \citenamefont {Bacon}, \citenamefont {Zhu}, \citenamefont {Abdelrasoul},
  \citenamefont {Li}, \citenamefont {Heinz}, \citenamefont {da~Jornada},
  \citenamefont {Cao},\ and\ \citenamefont {Dani}}]{Man2021}%
  \BibitemOpen
  \bibfield  {author} {\bibinfo {author} {\bibfnamefont {M.~K.~L.}\
  \bibnamefont {Man}}, \bibinfo {author} {\bibfnamefont {J.}~\bibnamefont
  {Madéo}}, \bibinfo {author} {\bibfnamefont {C.}~\bibnamefont {Sahoo}},
  \bibinfo {author} {\bibfnamefont {K.}~\bibnamefont {Xie}}, \bibinfo {author}
  {\bibfnamefont {M.}~\bibnamefont {Campbell}}, \bibinfo {author}
  {\bibfnamefont {V.}~\bibnamefont {Pareek}}, \bibinfo {author} {\bibfnamefont
  {A.}~\bibnamefont {Karmakar}}, \bibinfo {author} {\bibfnamefont {E.~L.}\
  \bibnamefont {Wong}}, \bibinfo {author} {\bibfnamefont {A.}~\bibnamefont
  {Al-Mahboob}}, \bibinfo {author} {\bibfnamefont {N.~S.}\ \bibnamefont
  {Chan}}, \bibinfo {author} {\bibfnamefont {D.~R.}\ \bibnamefont {Bacon}},
  \bibinfo {author} {\bibfnamefont {X.}~\bibnamefont {Zhu}}, \bibinfo {author}
  {\bibfnamefont {M.~M.~M.}\ \bibnamefont {Abdelrasoul}}, \bibinfo {author}
  {\bibfnamefont {X.}~\bibnamefont {Li}}, \bibinfo {author} {\bibfnamefont
  {T.~F.}\ \bibnamefont {Heinz}}, \bibinfo {author} {\bibfnamefont {F.~H.}\
  \bibnamefont {da~Jornada}}, \bibinfo {author} {\bibfnamefont
  {T.}~\bibnamefont {Cao}},\ and\ \bibinfo {author} {\bibfnamefont {K.~M.}\
  \bibnamefont {Dani}},\ }\href {https://doi.org/10.1126/sciadv.abg0192}
  {\bibfield  {journal} {\bibinfo  {journal} {Science Advances}\ }\textbf
  {\bibinfo {volume} {7}},\ \bibinfo {pages} {eabg0192} (\bibinfo {year}
  {2021})},\ \Eprint
  {https://arxiv.org/abs/https://www.science.org/doi/pdf/10.1126/sciadv.abg0192}
  {https://www.science.org/doi/pdf/10.1126/sciadv.abg0192} \BibitemShut
  {NoStop}%
\bibitem [{\citenamefont {Dong}\ \emph {et~al.}(2021)\citenamefont {Dong},
  \citenamefont {Puppin}, \citenamefont {Pincelli}, \citenamefont {Beaulieu},
  \citenamefont {Christiansen}, \citenamefont {Hübener}, \citenamefont
  {Nicholson}, \citenamefont {Xian}, \citenamefont {Dendzik}, \citenamefont
  {Deng}, \citenamefont {Windsor}, \citenamefont {Selig}, \citenamefont
  {Malic}, \citenamefont {Rubio}, \citenamefont {Knorr}, \citenamefont {Wolf},
  \citenamefont {Rettig},\ and\ \citenamefont {Ernstorfer}}]{Dong2021}%
  \BibitemOpen
  \bibfield  {author} {\bibinfo {author} {\bibfnamefont {S.}~\bibnamefont
  {Dong}}, \bibinfo {author} {\bibfnamefont {M.}~\bibnamefont {Puppin}},
  \bibinfo {author} {\bibfnamefont {T.}~\bibnamefont {Pincelli}}, \bibinfo
  {author} {\bibfnamefont {S.}~\bibnamefont {Beaulieu}}, \bibinfo {author}
  {\bibfnamefont {D.}~\bibnamefont {Christiansen}}, \bibinfo {author}
  {\bibfnamefont {H.}~\bibnamefont {Hübener}}, \bibinfo {author}
  {\bibfnamefont {C.}~\bibnamefont {Nicholson}}, \bibinfo {author}
  {\bibfnamefont {R.}~\bibnamefont {Xian}}, \bibinfo {author} {\bibfnamefont
  {M.}~\bibnamefont {Dendzik}}, \bibinfo {author} {\bibfnamefont
  {Y.}~\bibnamefont {Deng}}, \bibinfo {author} {\bibfnamefont {Y.}~\bibnamefont
  {Windsor}}, \bibinfo {author} {\bibfnamefont {M.}~\bibnamefont {Selig}},
  \bibinfo {author} {\bibfnamefont {E.}~\bibnamefont {Malic}}, \bibinfo
  {author} {\bibfnamefont {A.}~\bibnamefont {Rubio}}, \bibinfo {author}
  {\bibfnamefont {A.}~\bibnamefont {Knorr}}, \bibinfo {author} {\bibfnamefont
  {M.}~\bibnamefont {Wolf}}, \bibinfo {author} {\bibfnamefont {L.}~\bibnamefont
  {Rettig}},\ and\ \bibinfo {author} {\bibfnamefont {R.}~\bibnamefont
  {Ernstorfer}},\ }\href {https://doi.org/10.1002/ntls.10010} {\bibfield
  {journal} {\bibinfo  {journal} {Natural Sciences}\ }\textbf {\bibinfo
  {volume} {1}} (\bibinfo {year} {2021})}\BibitemShut {NoStop}%
\bibitem [{\citenamefont {Wallauer}\ \emph
  {et~al.}(2021{\natexlab{a}})\citenamefont {Wallauer}, \citenamefont {Raths},
  \citenamefont {Stallberg}, \citenamefont {M\"unster}, \citenamefont
  {Brandstetter}, \citenamefont {Yang}, \citenamefont {G\"udde}, \citenamefont
  {Puschnig}, \citenamefont {Soubatch}, \citenamefont {Kumpf}, \citenamefont
  {Bocquet}, \citenamefont {Tautz},\ and\ \citenamefont
  {H\"ofer}}]{Wallauer2021}%
  \BibitemOpen
  \bibfield  {author} {\bibinfo {author} {\bibfnamefont {R.}~\bibnamefont
  {Wallauer}}, \bibinfo {author} {\bibfnamefont {M.}~\bibnamefont {Raths}},
  \bibinfo {author} {\bibfnamefont {K.}~\bibnamefont {Stallberg}}, \bibinfo
  {author} {\bibfnamefont {L.}~\bibnamefont {M\"unster}}, \bibinfo {author}
  {\bibfnamefont {D.}~\bibnamefont {Brandstetter}}, \bibinfo {author}
  {\bibfnamefont {X.}~\bibnamefont {Yang}}, \bibinfo {author} {\bibfnamefont
  {J.}~\bibnamefont {G\"udde}}, \bibinfo {author} {\bibfnamefont
  {P.}~\bibnamefont {Puschnig}}, \bibinfo {author} {\bibfnamefont
  {S.}~\bibnamefont {Soubatch}}, \bibinfo {author} {\bibfnamefont
  {C.}~\bibnamefont {Kumpf}}, \bibinfo {author} {\bibfnamefont {F.~C.}\
  \bibnamefont {Bocquet}}, \bibinfo {author} {\bibfnamefont {F.~S.}\
  \bibnamefont {Tautz}},\ and\ \bibinfo {author} {\bibfnamefont
  {U.}~\bibnamefont {H\"ofer}},\ }\href
  {https://doi.org/10.1126/science.abf3286} {\bibfield  {journal} {\bibinfo
  {journal} {Science}\ }\textbf {\bibinfo {volume} {371}},\ \bibinfo {pages}
  {1056} (\bibinfo {year} {2021}{\natexlab{a}})}\BibitemShut {NoStop}%
\bibitem [{\citenamefont {Wallauer}\ \emph
  {et~al.}(2021{\natexlab{b}})\citenamefont {Wallauer}, \citenamefont
  {Perea-Causin}, \citenamefont {M\"unster}, \citenamefont {Zajusch},
  \citenamefont {Brem}, \citenamefont {G\"udde}, \citenamefont {Tanimura},
  \citenamefont {Lin}, \citenamefont {Huber}, \citenamefont {Malic},\ and\
  \citenamefont {H\"ofer}}]{Wallauer2021a}%
  \BibitemOpen
  \bibfield  {author} {\bibinfo {author} {\bibfnamefont {R.}~\bibnamefont
  {Wallauer}}, \bibinfo {author} {\bibfnamefont {R.}~\bibnamefont
  {Perea-Causin}}, \bibinfo {author} {\bibfnamefont {L.}~\bibnamefont
  {M\"unster}}, \bibinfo {author} {\bibfnamefont {S.}~\bibnamefont {Zajusch}},
  \bibinfo {author} {\bibfnamefont {S.}~\bibnamefont {Brem}}, \bibinfo {author}
  {\bibfnamefont {J.}~\bibnamefont {G\"udde}}, \bibinfo {author} {\bibfnamefont
  {K.}~\bibnamefont {Tanimura}}, \bibinfo {author} {\bibfnamefont {K.-Q.}\
  \bibnamefont {Lin}}, \bibinfo {author} {\bibfnamefont {R.}~\bibnamefont
  {Huber}}, \bibinfo {author} {\bibfnamefont {E.}~\bibnamefont {Malic}},\ and\
  \bibinfo {author} {\bibfnamefont {U.}~\bibnamefont {H\"ofer}},\ }\href
  {https://doi.org/10.1021/acs.nanolett.1c01839} {\bibfield  {journal}
  {\bibinfo  {journal} {Nano Letters}\ }\textbf {\bibinfo {volume} {xxx}},\
  \bibinfo {pages} {xxx} (\bibinfo {year} {2021}{\natexlab{b}})}\BibitemShut
  {NoStop}%
\bibitem [{\citenamefont {Bennecke}\ \emph {et~al.}(2024)\citenamefont
  {Bennecke}, \citenamefont {Windischbacher}, \citenamefont {Schmitt},
  \citenamefont {Bange}, \citenamefont {Hemm}, \citenamefont {Kern},
  \citenamefont {D'Avino}, \citenamefont {Blase}, \citenamefont {Steil},
  \citenamefont {Steil}, \citenamefont {Aeschlimann}, \citenamefont
  {Stadtm\"uller}, \citenamefont {Reutzel}, \citenamefont {Puschnig},
  \citenamefont {Jansen},\ and\ \citenamefont {Mathias}}]{Bennecke2024}%
  \BibitemOpen
  \bibfield  {author} {\bibinfo {author} {\bibfnamefont {W.}~\bibnamefont
  {Bennecke}}, \bibinfo {author} {\bibfnamefont {A.}~\bibnamefont
  {Windischbacher}}, \bibinfo {author} {\bibfnamefont {D.}~\bibnamefont
  {Schmitt}}, \bibinfo {author} {\bibfnamefont {J.~P.}\ \bibnamefont {Bange}},
  \bibinfo {author} {\bibfnamefont {R.}~\bibnamefont {Hemm}}, \bibinfo {author}
  {\bibfnamefont {C.~S.}\ \bibnamefont {Kern}}, \bibinfo {author}
  {\bibfnamefont {G.}~\bibnamefont {D'Avino}}, \bibinfo {author} {\bibfnamefont
  {X.}~\bibnamefont {Blase}}, \bibinfo {author} {\bibfnamefont
  {D.}~\bibnamefont {Steil}}, \bibinfo {author} {\bibfnamefont
  {S.}~\bibnamefont {Steil}}, \bibinfo {author} {\bibfnamefont
  {M.}~\bibnamefont {Aeschlimann}}, \bibinfo {author} {\bibfnamefont
  {B.}~\bibnamefont {Stadtm\"uller}}, \bibinfo {author} {\bibfnamefont
  {M.}~\bibnamefont {Reutzel}}, \bibinfo {author} {\bibfnamefont
  {P.}~\bibnamefont {Puschnig}}, \bibinfo {author} {\bibfnamefont {G.~S.~M.}\
  \bibnamefont {Jansen}},\ and\ \bibinfo {author} {\bibfnamefont
  {S.}~\bibnamefont {Mathias}},\ }\href
  {https://doi.org/10.1038/s41467-024-45973-x} {\bibfield  {journal} {\bibinfo
  {journal} {Nature Communications}\ }\textbf {\bibinfo {volume} {15}},\
  \bibinfo {pages} {1804} (\bibinfo {year} {2024})}\BibitemShut {NoStop}%
\bibitem [{\citenamefont {Reutzel}\ \emph {et~al.}(2024)\citenamefont
  {Reutzel}, \citenamefont {Jansen},\ and\ \citenamefont
  {Mathias}}]{Reutzel2024}%
  \BibitemOpen
  \bibfield  {author} {\bibinfo {author} {\bibfnamefont {M.}~\bibnamefont
  {Reutzel}}, \bibinfo {author} {\bibfnamefont {G.~S.~M.}\ \bibnamefont
  {Jansen}},\ and\ \bibinfo {author} {\bibfnamefont {S.}~\bibnamefont
  {Mathias}},\ }\href {https://doi.org/10.1080/23746149.2024.2378722}
  {\bibfield  {journal} {\bibinfo  {journal} {Advances in Physics: X}\ }\textbf
  {\bibinfo {volume} {9}},\ \bibinfo {pages} {2378722} (\bibinfo {year}
  {2024})}\BibitemShut {NoStop}%
\bibitem [{\citenamefont {Dauth}\ and\ \citenamefont
  {K\"ummel}(2016)}]{Dauth2016b}%
  \BibitemOpen
  \bibfield  {author} {\bibinfo {author} {\bibfnamefont {M.}~\bibnamefont
  {Dauth}}\ and\ \bibinfo {author} {\bibfnamefont {S.}~\bibnamefont
  {K\"ummel}},\ }\href {https://doi.org/10.1103/PhysRevA.93.022502} {\bibfield
  {journal} {\bibinfo  {journal} {Phys. Rev. A}\ }\textbf {\bibinfo {volume}
  {93}},\ \bibinfo {pages} {022502} (\bibinfo {year} {2016})}\BibitemShut
  {NoStop}%
\bibitem [{\citenamefont {De~Giovannini}\ \emph {et~al.}(2017)\citenamefont
  {De~Giovannini}, \citenamefont {Hübener},\ and\ \citenamefont
  {Rubio}}]{DeGiovannini2017}%
  \BibitemOpen
  \bibfield  {author} {\bibinfo {author} {\bibfnamefont {U.}~\bibnamefont
  {De~Giovannini}}, \bibinfo {author} {\bibfnamefont {H.}~\bibnamefont
  {Hübener}},\ and\ \bibinfo {author} {\bibfnamefont {A.}~\bibnamefont
  {Rubio}},\ }\href {https://doi.org/10.1021/acs.jctc.6b00897} {\bibfield
  {journal} {\bibinfo  {journal} {J. Chem. Theory Comput.}\ }\textbf {\bibinfo
  {volume} {13}},\ \bibinfo {pages} {265} (\bibinfo {year} {2017})}\BibitemShut
  {NoStop}%
\bibitem [{\citenamefont {Dauth}\ \emph {et~al.}(2016)\citenamefont {Dauth},
  \citenamefont {Graus}, \citenamefont {Schelter}, \citenamefont {Wie\ss{}ner},
  \citenamefont {Sch\"oll}, \citenamefont {Reinert},\ and\ \citenamefont
  {K\"ummel}}]{Dauth2016a}%
  \BibitemOpen
  \bibfield  {author} {\bibinfo {author} {\bibfnamefont {M.}~\bibnamefont
  {Dauth}}, \bibinfo {author} {\bibfnamefont {M.}~\bibnamefont {Graus}},
  \bibinfo {author} {\bibfnamefont {I.}~\bibnamefont {Schelter}}, \bibinfo
  {author} {\bibfnamefont {M.}~\bibnamefont {Wie\ss{}ner}}, \bibinfo {author}
  {\bibfnamefont {A.}~\bibnamefont {Sch\"oll}}, \bibinfo {author}
  {\bibfnamefont {F.}~\bibnamefont {Reinert}},\ and\ \bibinfo {author}
  {\bibfnamefont {S.}~\bibnamefont {K\"ummel}},\ }\href
  {https://doi.org/10.1103/PhysRevLett.117.183001} {\bibfield  {journal}
  {\bibinfo  {journal} {Phys. Rev. Lett.}\ }\textbf {\bibinfo {volume} {117}},\
  \bibinfo {pages} {183001} (\bibinfo {year} {2016})}\BibitemShut {NoStop}%
\bibitem [{\citenamefont {Kern}\ \emph
  {et~al.}(2023{\natexlab{a}})\citenamefont {Kern}, \citenamefont {Haags},
  \citenamefont {Egger}, \citenamefont {Yang}, \citenamefont {Kirschner},
  \citenamefont {Wolff}, \citenamefont {Seyller}, \citenamefont {Gottwald},
  \citenamefont {Richter}, \citenamefont {De~Giovannini}, \citenamefont
  {Rubio}, \citenamefont {Ramsey}, \citenamefont {Bocquet}, \citenamefont
  {Soubatch}, \citenamefont {Tautz}, \citenamefont {Puschnig},\ and\
  \citenamefont {Moser}}]{Kern2023}%
  \BibitemOpen
  \bibfield  {author} {\bibinfo {author} {\bibfnamefont {C.~S.}\ \bibnamefont
  {Kern}}, \bibinfo {author} {\bibfnamefont {A.}~\bibnamefont {Haags}},
  \bibinfo {author} {\bibfnamefont {L.}~\bibnamefont {Egger}}, \bibinfo
  {author} {\bibfnamefont {X.}~\bibnamefont {Yang}}, \bibinfo {author}
  {\bibfnamefont {H.}~\bibnamefont {Kirschner}}, \bibinfo {author}
  {\bibfnamefont {S.}~\bibnamefont {Wolff}}, \bibinfo {author} {\bibfnamefont
  {T.}~\bibnamefont {Seyller}}, \bibinfo {author} {\bibfnamefont
  {A.}~\bibnamefont {Gottwald}}, \bibinfo {author} {\bibfnamefont
  {M.}~\bibnamefont {Richter}}, \bibinfo {author} {\bibfnamefont
  {U.}~\bibnamefont {De~Giovannini}}, \bibinfo {author} {\bibfnamefont
  {A.}~\bibnamefont {Rubio}}, \bibinfo {author} {\bibfnamefont {M.~G.}\
  \bibnamefont {Ramsey}}, \bibinfo {author} {\bibfnamefont {F.~C.}\
  \bibnamefont {Bocquet}}, \bibinfo {author} {\bibfnamefont {S.}~\bibnamefont
  {Soubatch}}, \bibinfo {author} {\bibfnamefont {F.~S.}\ \bibnamefont {Tautz}},
  \bibinfo {author} {\bibfnamefont {P.}~\bibnamefont {Puschnig}},\ and\
  \bibinfo {author} {\bibfnamefont {S.}~\bibnamefont {Moser}},\ }\href
  {https://doi.org/10.1103/PhysRevResearch.5.033075} {\bibfield  {journal}
  {\bibinfo  {journal} {Phys. Rev. Research}\ }\textbf {\bibinfo {volume}
  {5}},\ \bibinfo {pages} {033075} (\bibinfo {year}
  {2023}{\natexlab{a}})}\BibitemShut {NoStop}%
\bibitem [{\citenamefont {Stefanucci}\ and\ \citenamefont {van
  Leeuwen}(2025)}]{Stefanucci2025}%
  \BibitemOpen
  \bibfield  {author} {\bibinfo {author} {\bibfnamefont {G.}~\bibnamefont
  {Stefanucci}}\ and\ \bibinfo {author} {\bibfnamefont {R.}~\bibnamefont {van
  Leeuwen}},\ }\href@noop {} {\emph {\bibinfo {title} {Nonequilibrium Many-Body
  Theory of Quantum Systems: A Modern Introduction}}},\ \bibinfo {edition}
  {2nd}\ ed.\ (\bibinfo  {publisher} {Cambridge University Press},\ \bibinfo
  {year} {2025})\BibitemShut {NoStop}%
\bibitem [{\citenamefont {Freericks}\ \emph {et~al.}(2009)\citenamefont
  {Freericks}, \citenamefont {Krishnamurthy},\ and\ \citenamefont
  {Pruschke}}]{Freericks2009}%
  \BibitemOpen
  \bibfield  {author} {\bibinfo {author} {\bibfnamefont {J.~K.}\ \bibnamefont
  {Freericks}}, \bibinfo {author} {\bibfnamefont {H.~R.}\ \bibnamefont
  {Krishnamurthy}},\ and\ \bibinfo {author} {\bibfnamefont {T.}~\bibnamefont
  {Pruschke}},\ }\href {https://doi.org/10.1103/PhysRevLett.102.136401}
  {\bibfield  {journal} {\bibinfo  {journal} {Phys. Rev. Lett.}\ }\textbf
  {\bibinfo {volume} {102}},\ \bibinfo {pages} {136401} (\bibinfo {year}
  {2009})}\BibitemShut {NoStop}%
\bibitem [{\citenamefont {Perfetto}\ \emph {et~al.}(2016)\citenamefont
  {Perfetto}, \citenamefont {Sangalli}, \citenamefont {Marini},\ and\
  \citenamefont {Stefanucci}}]{Perfetto2016}%
  \BibitemOpen
  \bibfield  {author} {\bibinfo {author} {\bibfnamefont {E.}~\bibnamefont
  {Perfetto}}, \bibinfo {author} {\bibfnamefont {D.}~\bibnamefont {Sangalli}},
  \bibinfo {author} {\bibfnamefont {A.}~\bibnamefont {Marini}},\ and\ \bibinfo
  {author} {\bibfnamefont {G.}~\bibnamefont {Stefanucci}},\ }\href
  {https://doi.org/10.1103/PhysRevB.94.245303} {\bibfield  {journal} {\bibinfo
  {journal} {Phys. Rev. B}\ }\textbf {\bibinfo {volume} {94}},\ \bibinfo
  {pages} {245303} (\bibinfo {year} {2016})}\BibitemShut {NoStop}%
\bibitem [{\citenamefont {Sch{\"u}ler}\ and\ \citenamefont
  {Sentef}(2021)}]{Schueler2021}%
  \BibitemOpen
  \bibfield  {author} {\bibinfo {author} {\bibfnamefont {M.}~\bibnamefont
  {Sch{\"u}ler}}\ and\ \bibinfo {author} {\bibfnamefont {M.~A.}\ \bibnamefont
  {Sentef}},\ }\href
  {https://doi.org/https://doi.org/10.1016/j.elspec.2021.147121} {\bibfield
  {journal} {\bibinfo  {journal} {Journal of Electron Spectroscopy and Related
  Phenomena}\ }\textbf {\bibinfo {volume} {253}},\ \bibinfo {pages} {147121}
  (\bibinfo {year} {2021})}\BibitemShut {NoStop}%
\bibitem [{\citenamefont {Rustagi}\ and\ \citenamefont
  {Kemper}(2018)}]{Rustagi2018}%
  \BibitemOpen
  \bibfield  {author} {\bibinfo {author} {\bibfnamefont {A.}~\bibnamefont
  {Rustagi}}\ and\ \bibinfo {author} {\bibfnamefont {A.~F.}\ \bibnamefont
  {Kemper}},\ }\href {https://doi.org/10.1103/PhysRevB.97.235310} {\bibfield
  {journal} {\bibinfo  {journal} {Phys. Rev. B}\ }\textbf {\bibinfo {volume}
  {97}},\ \bibinfo {pages} {235310} (\bibinfo {year} {2018})}\BibitemShut
  {NoStop}%
\bibitem [{\citenamefont {Sangalli}(2021)}]{Sangalli2021}%
  \BibitemOpen
  \bibfield  {author} {\bibinfo {author} {\bibfnamefont {D.}~\bibnamefont
  {Sangalli}},\ }\href {https://doi.org/10.1103/PhysRevMaterials.5.083803}
  {\bibfield  {journal} {\bibinfo  {journal} {Phys. Rev. Mater.}\ }\textbf
  {\bibinfo {volume} {5}},\ \bibinfo {pages} {083803} (\bibinfo {year}
  {2021})}\BibitemShut {NoStop}%
\bibitem [{\citenamefont {T\"orm\"a}(2023)}]{Torma2023}%
  \BibitemOpen
  \bibfield  {author} {\bibinfo {author} {\bibfnamefont {P.}~\bibnamefont
  {T\"orm\"a}},\ }\href {https://doi.org/10.1103/PhysRevLett.131.240001}
  {\bibfield  {journal} {\bibinfo  {journal} {Phys. Rev. Lett.}\ }\textbf
  {\bibinfo {volume} {131}},\ \bibinfo {pages} {240001} (\bibinfo {year}
  {2023})}\BibitemShut {NoStop}%
\bibitem [{\citenamefont {Kang}\ \emph {et~al.}(2025)\citenamefont {Kang},
  \citenamefont {Kim}, \citenamefont {Qian}, \citenamefont {Neves},
  \citenamefont {Ye}, \citenamefont {Jung}, \citenamefont {Puntel},
  \citenamefont {Mazzola}, \citenamefont {Fang}, \citenamefont {Jozwiak},
  \citenamefont {Bostwick}, \citenamefont {Rotenberg}, \citenamefont {Fuji},
  \citenamefont {Vobornik}, \citenamefont {Park}, \citenamefont {Checkelsky},
  \citenamefont {Yang},\ and\ \citenamefont {Comin}}]{Kang2025}%
  \BibitemOpen
  \bibfield  {author} {\bibinfo {author} {\bibfnamefont {M.}~\bibnamefont
  {Kang}}, \bibinfo {author} {\bibfnamefont {S.}~\bibnamefont {Kim}}, \bibinfo
  {author} {\bibfnamefont {Y.}~\bibnamefont {Qian}}, \bibinfo {author}
  {\bibfnamefont {P.~M.}\ \bibnamefont {Neves}}, \bibinfo {author}
  {\bibfnamefont {L.}~\bibnamefont {Ye}}, \bibinfo {author} {\bibfnamefont
  {J.}~\bibnamefont {Jung}}, \bibinfo {author} {\bibfnamefont {D.}~\bibnamefont
  {Puntel}}, \bibinfo {author} {\bibfnamefont {F.}~\bibnamefont {Mazzola}},
  \bibinfo {author} {\bibfnamefont {S.}~\bibnamefont {Fang}}, \bibinfo {author}
  {\bibfnamefont {C.}~\bibnamefont {Jozwiak}}, \bibinfo {author} {\bibfnamefont
  {A.}~\bibnamefont {Bostwick}}, \bibinfo {author} {\bibfnamefont
  {E.}~\bibnamefont {Rotenberg}}, \bibinfo {author} {\bibfnamefont
  {J.}~\bibnamefont {Fuji}}, \bibinfo {author} {\bibfnamefont {I.}~\bibnamefont
  {Vobornik}}, \bibinfo {author} {\bibfnamefont {J.-H.}\ \bibnamefont {Park}},
  \bibinfo {author} {\bibfnamefont {J.~G.}\ \bibnamefont {Checkelsky}},
  \bibinfo {author} {\bibfnamefont {B.-J.}\ \bibnamefont {Yang}},\ and\
  \bibinfo {author} {\bibfnamefont {R.}~\bibnamefont {Comin}},\ }\href
  {https://doi.org/10.1038/s41567-024-02678-8} {\bibfield  {journal} {\bibinfo
  {journal} {Nature Physics}\ }\textbf {\bibinfo {volume} {21}},\ \bibinfo
  {pages} {110} (\bibinfo {year} {2025})}\BibitemShut {NoStop}%
\bibitem [{\citenamefont {Cho}\ \emph {et~al.}(2018)\citenamefont {Cho},
  \citenamefont {Park}, \citenamefont {Hong}, \citenamefont {Jung},
  \citenamefont {Kim}, \citenamefont {Han}, \citenamefont {Kyung},
  \citenamefont {Kim}, \citenamefont {Mo}, \citenamefont {Denlinger},
  \citenamefont {Shim}, \citenamefont {Han}, \citenamefont {Kim},\ and\
  \citenamefont {Park}}]{Cho2018}%
  \BibitemOpen
  \bibfield  {author} {\bibinfo {author} {\bibfnamefont {S.}~\bibnamefont
  {Cho}}, \bibinfo {author} {\bibfnamefont {J.-H.}\ \bibnamefont {Park}},
  \bibinfo {author} {\bibfnamefont {J.}~\bibnamefont {Hong}}, \bibinfo {author}
  {\bibfnamefont {J.}~\bibnamefont {Jung}}, \bibinfo {author} {\bibfnamefont
  {B.~S.}\ \bibnamefont {Kim}}, \bibinfo {author} {\bibfnamefont
  {G.}~\bibnamefont {Han}}, \bibinfo {author} {\bibfnamefont {W.}~\bibnamefont
  {Kyung}}, \bibinfo {author} {\bibfnamefont {Y.}~\bibnamefont {Kim}}, \bibinfo
  {author} {\bibfnamefont {S.-K.}\ \bibnamefont {Mo}}, \bibinfo {author}
  {\bibfnamefont {J.~D.}\ \bibnamefont {Denlinger}}, \bibinfo {author}
  {\bibfnamefont {J.~H.}\ \bibnamefont {Shim}}, \bibinfo {author}
  {\bibfnamefont {J.~H.}\ \bibnamefont {Han}}, \bibinfo {author} {\bibfnamefont
  {C.}~\bibnamefont {Kim}},\ and\ \bibinfo {author} {\bibfnamefont {S.~R.}\
  \bibnamefont {Park}},\ }\href
  {https://doi.org/10.1103/PhysRevLett.121.186401} {\bibfield  {journal}
  {\bibinfo  {journal} {Phys. Rev. Lett.}\ }\textbf {\bibinfo {volume} {121}},\
  \bibinfo {pages} {186401} (\bibinfo {year} {2018})}\BibitemShut {NoStop}%
\bibitem [{\citenamefont {Sch\"uler}\ \emph
  {et~al.}(2020{\natexlab{a}})\citenamefont {Sch\"uler}, \citenamefont
  {Giovannini}, \citenamefont {H\"ubener}, \citenamefont {Rubio}, \citenamefont
  {Sentef},\ and\ \citenamefont {Werner}}]{Schueler2020}%
  \BibitemOpen
  \bibfield  {author} {\bibinfo {author} {\bibfnamefont {M.}~\bibnamefont
  {Sch\"uler}}, \bibinfo {author} {\bibfnamefont {U.~D.}\ \bibnamefont
  {Giovannini}}, \bibinfo {author} {\bibfnamefont {H.}~\bibnamefont
  {H\"ubener}}, \bibinfo {author} {\bibfnamefont {A.}~\bibnamefont {Rubio}},
  \bibinfo {author} {\bibfnamefont {M.~A.}\ \bibnamefont {Sentef}},\ and\
  \bibinfo {author} {\bibfnamefont {P.}~\bibnamefont {Werner}},\ }\href
  {https://doi.org/10.1126/sciadv.aay2730} {\bibfield  {journal} {\bibinfo
  {journal} {Science Advances}\ }\textbf {\bibinfo {volume} {6}},\ \bibinfo
  {pages} {eaay2730} (\bibinfo {year} {2020}{\natexlab{a}})}\BibitemShut
  {NoStop}%
\bibitem [{\citenamefont {{\"U}nzelmann}\ \emph {et~al.}(2021)\citenamefont
  {{\"U}nzelmann}, \citenamefont {Bentmann}, \citenamefont {Figgemeier},
  \citenamefont {Eck}, \citenamefont {Neu}, \citenamefont {Geldiyev},
  \citenamefont {Diekmann}, \citenamefont {Rohlf}, \citenamefont {Buck},
  \citenamefont {Hoesch}, \citenamefont {Kall{\"a}ne}, \citenamefont
  {Rossnagel}, \citenamefont {Thomale}, \citenamefont {Siegrist}, \citenamefont
  {Sangiovanni}, \citenamefont {Sante},\ and\ \citenamefont
  {Reinert}}]{Unzelmann2021}%
  \BibitemOpen
  \bibfield  {author} {\bibinfo {author} {\bibfnamefont {M.}~\bibnamefont
  {{\"U}nzelmann}}, \bibinfo {author} {\bibfnamefont {H.}~\bibnamefont
  {Bentmann}}, \bibinfo {author} {\bibfnamefont {T.}~\bibnamefont
  {Figgemeier}}, \bibinfo {author} {\bibfnamefont {P.}~\bibnamefont {Eck}},
  \bibinfo {author} {\bibfnamefont {J.~N.}\ \bibnamefont {Neu}}, \bibinfo
  {author} {\bibfnamefont {B.}~\bibnamefont {Geldiyev}}, \bibinfo {author}
  {\bibfnamefont {F.}~\bibnamefont {Diekmann}}, \bibinfo {author}
  {\bibfnamefont {S.}~\bibnamefont {Rohlf}}, \bibinfo {author} {\bibfnamefont
  {J.}~\bibnamefont {Buck}}, \bibinfo {author} {\bibfnamefont {M.}~\bibnamefont
  {Hoesch}}, \bibinfo {author} {\bibfnamefont {M.}~\bibnamefont {Kall{\"a}ne}},
  \bibinfo {author} {\bibfnamefont {K.}~\bibnamefont {Rossnagel}}, \bibinfo
  {author} {\bibfnamefont {R.}~\bibnamefont {Thomale}}, \bibinfo {author}
  {\bibfnamefont {T.}~\bibnamefont {Siegrist}}, \bibinfo {author}
  {\bibfnamefont {G.}~\bibnamefont {Sangiovanni}}, \bibinfo {author}
  {\bibfnamefont {D.~D.}\ \bibnamefont {Sante}},\ and\ \bibinfo {author}
  {\bibfnamefont {F.}~\bibnamefont {Reinert}},\ }\href
  {https://doi.org/10.1038/s41467-021-23727-3} {\bibfield  {journal} {\bibinfo
  {journal} {Nature Communications}\ }\textbf {\bibinfo {volume} {12}},\
  \bibinfo {pages} {3650} (\bibinfo {year} {2021})}\BibitemShut {NoStop}%
\bibitem [{\citenamefont {Sch\"uler}\ \emph
  {et~al.}(2020{\natexlab{b}})\citenamefont {Sch\"uler}, \citenamefont
  {De~Giovannini}, \citenamefont {H\"ubener}, \citenamefont {Rubio},
  \citenamefont {Sentef}, \citenamefont {Devereaux},\ and\ \citenamefont
  {Werner}}]{Schueler2020a}%
  \BibitemOpen
  \bibfield  {author} {\bibinfo {author} {\bibfnamefont {M.}~\bibnamefont
  {Sch\"uler}}, \bibinfo {author} {\bibfnamefont {U.}~\bibnamefont
  {De~Giovannini}}, \bibinfo {author} {\bibfnamefont {H.}~\bibnamefont
  {H\"ubener}}, \bibinfo {author} {\bibfnamefont {A.}~\bibnamefont {Rubio}},
  \bibinfo {author} {\bibfnamefont {M.~A.}\ \bibnamefont {Sentef}}, \bibinfo
  {author} {\bibfnamefont {T.~P.}\ \bibnamefont {Devereaux}},\ and\ \bibinfo
  {author} {\bibfnamefont {P.}~\bibnamefont {Werner}},\ }\href
  {https://doi.org/10.1103/PhysRevX.10.041013} {\bibfield  {journal} {\bibinfo
  {journal} {Phys. Rev. X}\ }\textbf {\bibinfo {volume} {10}},\ \bibinfo
  {pages} {041013} (\bibinfo {year} {2020}{\natexlab{b}})}\BibitemShut
  {NoStop}%
\bibitem [{\citenamefont {Adawi}(1964)}]{Adawi1964}%
  \BibitemOpen
  \bibfield  {author} {\bibinfo {author} {\bibfnamefont {I.}~\bibnamefont
  {Adawi}},\ }\href {https://doi.org/10.1103/PhysRev.134.A788} {\bibfield
  {journal} {\bibinfo  {journal} {Phys. Rev.}\ }\textbf {\bibinfo {volume}
  {134}},\ \bibinfo {pages} {A788} (\bibinfo {year} {1964})}\BibitemShut
  {NoStop}%
\bibitem [{\citenamefont {Feibelman}\ and\ \citenamefont
  {Eastman}(1974)}]{Feibelman1974}%
  \BibitemOpen
  \bibfield  {author} {\bibinfo {author} {\bibfnamefont {P.~J.}\ \bibnamefont
  {Feibelman}}\ and\ \bibinfo {author} {\bibfnamefont {D.~E.}\ \bibnamefont
  {Eastman}},\ }\href {https://doi.org/10.1103/PhysRevB.10.4932} {\bibfield
  {journal} {\bibinfo  {journal} {Phys. Rev. B}\ }\textbf {\bibinfo {volume}
  {10}},\ \bibinfo {pages} {4932} (\bibinfo {year} {1974})}\BibitemShut
  {NoStop}%
\bibitem [{\citenamefont {Puschnig}\ \emph {et~al.}(2009)\citenamefont
  {Puschnig}, \citenamefont {Berkebile}, \citenamefont {Fleming}, \citenamefont
  {Koller}, \citenamefont {Emtsev}, \citenamefont {Seyller}, \citenamefont
  {Riley}, \citenamefont {Ambrosch-Draxl}, \citenamefont {Netzer},\ and\
  \citenamefont {Ramsey}}]{Puschnig2009a}%
  \BibitemOpen
  \bibfield  {author} {\bibinfo {author} {\bibfnamefont {P.}~\bibnamefont
  {Puschnig}}, \bibinfo {author} {\bibfnamefont {S.}~\bibnamefont {Berkebile}},
  \bibinfo {author} {\bibfnamefont {A.~J.}\ \bibnamefont {Fleming}}, \bibinfo
  {author} {\bibfnamefont {G.}~\bibnamefont {Koller}}, \bibinfo {author}
  {\bibfnamefont {K.}~\bibnamefont {Emtsev}}, \bibinfo {author} {\bibfnamefont
  {T.}~\bibnamefont {Seyller}}, \bibinfo {author} {\bibfnamefont {J.~D.}\
  \bibnamefont {Riley}}, \bibinfo {author} {\bibfnamefont {C.}~\bibnamefont
  {Ambrosch-Draxl}}, \bibinfo {author} {\bibfnamefont {F.~P.}\ \bibnamefont
  {Netzer}},\ and\ \bibinfo {author} {\bibfnamefont {M.~G.}\ \bibnamefont
  {Ramsey}},\ }\href {https://doi.org/10.1126/science.1176105} {\bibfield
  {journal} {\bibinfo  {journal} {Science}\ }\textbf {\bibinfo {volume}
  {326}},\ \bibinfo {pages} {702} (\bibinfo {year} {2009})}\BibitemShut
  {NoStop}%
\bibitem [{\citenamefont {Kern}\ \emph
  {et~al.}(2023{\natexlab{b}})\citenamefont {Kern}, \citenamefont
  {Windischbacher},\ and\ \citenamefont {Puschnig}}]{Kern2023a}%
  \BibitemOpen
  \bibfield  {author} {\bibinfo {author} {\bibfnamefont {C.~S.}\ \bibnamefont
  {Kern}}, \bibinfo {author} {\bibfnamefont {A.}~\bibnamefont
  {Windischbacher}},\ and\ \bibinfo {author} {\bibfnamefont {P.}~\bibnamefont
  {Puschnig}},\ }\href {https://doi.org/10.1103/PhysRevB.108.085132} {\bibfield
   {journal} {\bibinfo  {journal} {Phys. Rev. B}\ }\textbf {\bibinfo {volume}
  {108}},\ \bibinfo {pages} {085132} (\bibinfo {year}
  {2023}{\natexlab{b}})}\BibitemShut {NoStop}%
\bibitem [{\citenamefont {Ye}\ \emph {et~al.}(2014)\citenamefont {Ye},
  \citenamefont {Cao}, \citenamefont {O'Brien}, \citenamefont {Zhu},
  \citenamefont {Yin}, \citenamefont {Wang}, \citenamefont {Louie},\ and\
  \citenamefont {Zhang}}]{Ye2014a}%
  \BibitemOpen
  \bibfield  {author} {\bibinfo {author} {\bibfnamefont {Z.}~\bibnamefont
  {Ye}}, \bibinfo {author} {\bibfnamefont {T.}~\bibnamefont {Cao}}, \bibinfo
  {author} {\bibfnamefont {K.}~\bibnamefont {O'Brien}}, \bibinfo {author}
  {\bibfnamefont {H.}~\bibnamefont {Zhu}}, \bibinfo {author} {\bibfnamefont
  {X.}~\bibnamefont {Yin}}, \bibinfo {author} {\bibfnamefont {Y.}~\bibnamefont
  {Wang}}, \bibinfo {author} {\bibfnamefont {S.~G.}\ \bibnamefont {Louie}},\
  and\ \bibinfo {author} {\bibfnamefont {X.}~\bibnamefont {Zhang}},\ }\href
  {https://doi.org/10.1038/nature13734} {\bibfield  {journal} {\bibinfo
  {journal} {Nature}\ }\textbf {\bibinfo {volume} {513}},\ \bibinfo {pages}
  {214} (\bibinfo {year} {2014})}\BibitemShut {NoStop}%
\bibitem [{\citenamefont {Wu}\ \emph {et~al.}(2015)\citenamefont {Wu},
  \citenamefont {Qu},\ and\ \citenamefont {MacDonald}}]{Wu2015}%
  \BibitemOpen
  \bibfield  {author} {\bibinfo {author} {\bibfnamefont {F.}~\bibnamefont
  {Wu}}, \bibinfo {author} {\bibfnamefont {F.}~\bibnamefont {Qu}},\ and\
  \bibinfo {author} {\bibfnamefont {A.~H.}\ \bibnamefont {MacDonald}},\ }\href
  {https://doi.org/10.1103/PhysRevB.91.075310} {\bibfield  {journal} {\bibinfo
  {journal} {Phys. Rev. B}\ }\textbf {\bibinfo {volume} {91}},\ \bibinfo
  {pages} {075310} (\bibinfo {year} {2015})}\BibitemShut {NoStop}%
\bibitem [{\citenamefont {Poellmann}\ \emph {et~al.}(2015)\citenamefont
  {Poellmann}, \citenamefont {Steinleitner}, \citenamefont {Leierseder},
  \citenamefont {Nagler}, \citenamefont {Plechinger}, \citenamefont {Porer},
  \citenamefont {Bratschitsch}, \citenamefont {Sch{\"u}ller}, \citenamefont
  {Korn},\ and\ \citenamefont {Huber}}]{Poellmann2015}%
  \BibitemOpen
  \bibfield  {author} {\bibinfo {author} {\bibfnamefont {C.}~\bibnamefont
  {Poellmann}}, \bibinfo {author} {\bibfnamefont {P.}~\bibnamefont
  {Steinleitner}}, \bibinfo {author} {\bibfnamefont {U.}~\bibnamefont
  {Leierseder}}, \bibinfo {author} {\bibfnamefont {P.}~\bibnamefont {Nagler}},
  \bibinfo {author} {\bibfnamefont {G.}~\bibnamefont {Plechinger}}, \bibinfo
  {author} {\bibfnamefont {M.}~\bibnamefont {Porer}}, \bibinfo {author}
  {\bibfnamefont {R.}~\bibnamefont {Bratschitsch}}, \bibinfo {author}
  {\bibfnamefont {C.}~\bibnamefont {Sch{\"u}ller}}, \bibinfo {author}
  {\bibfnamefont {T.}~\bibnamefont {Korn}},\ and\ \bibinfo {author}
  {\bibfnamefont {R.}~\bibnamefont {Huber}},\ }\href
  {https://doi.org/10.1038/nmat4356} {\bibfield  {journal} {\bibinfo  {journal}
  {Nature Materials}\ }\textbf {\bibinfo {volume} {14}},\ \bibinfo {pages}
  {889} (\bibinfo {year} {2015})}\BibitemShut {NoStop}%
\bibitem [{\citenamefont {Theilen}\ \emph {et~al.}(2025)\citenamefont
  {Theilen}, \citenamefont {Kaidisch}, \citenamefont {Stettner}, \citenamefont
  {Zajusch}, \citenamefont {Fackelman}, \citenamefont {Adamkiewicz},
  \citenamefont {Wallauer}, \citenamefont {Windischbacher}, \citenamefont
  {Kern}, \citenamefont {Ramsey}, \citenamefont {Bocquet}, \citenamefont
  {Soubatch}, \citenamefont {Tautz}, \citenamefont {Höfer},\ and\
  \citenamefont {Puschnig}}]{Theilen2025}%
  \BibitemOpen
  \bibfield  {author} {\bibinfo {author} {\bibfnamefont {M.}~\bibnamefont
  {Theilen}}, \bibinfo {author} {\bibfnamefont {S.}~\bibnamefont {Kaidisch}},
  \bibinfo {author} {\bibfnamefont {M.}~\bibnamefont {Stettner}}, \bibinfo
  {author} {\bibfnamefont {S.}~\bibnamefont {Zajusch}}, \bibinfo {author}
  {\bibfnamefont {E.}~\bibnamefont {Fackelman}}, \bibinfo {author}
  {\bibfnamefont {A.}~\bibnamefont {Adamkiewicz}}, \bibinfo {author}
  {\bibfnamefont {R.}~\bibnamefont {Wallauer}}, \bibinfo {author}
  {\bibfnamefont {A.}~\bibnamefont {Windischbacher}}, \bibinfo {author}
  {\bibfnamefont {C.~S.}\ \bibnamefont {Kern}}, \bibinfo {author}
  {\bibfnamefont {M.~G.}\ \bibnamefont {Ramsey}}, \bibinfo {author}
  {\bibfnamefont {F.~C.}\ \bibnamefont {Bocquet}}, \bibinfo {author}
  {\bibfnamefont {S.}~\bibnamefont {Soubatch}}, \bibinfo {author}
  {\bibfnamefont {F.~S.}\ \bibnamefont {Tautz}}, \bibinfo {author}
  {\bibfnamefont {U.}~\bibnamefont {Höfer}},\ and\ \bibinfo {author}
  {\bibfnamefont {P.}~\bibnamefont {Puschnig}},\ }\href
  {https://arxiv.org/abs/2511.23001} {\bibinfo {title} {Observing the spatial
  and temporal evolution of exciton wave functions}} (\bibinfo {year} {2025}),\
  \Eprint {https://arxiv.org/abs/2511.23001} {arXiv:2511.23001
  [cond-mat.mtrl-sci]} \BibitemShut {NoStop}%
\bibitem [{\citenamefont {Sangalli}\ \emph {et~al.}(2018)\citenamefont
  {Sangalli}, \citenamefont {Perfetto}, \citenamefont {Stefanucci},\ and\
  \citenamefont {Marini}}]{Sangalli2018}%
  \BibitemOpen
  \bibfield  {author} {\bibinfo {author} {\bibfnamefont {D.}~\bibnamefont
  {Sangalli}}, \bibinfo {author} {\bibfnamefont {E.}~\bibnamefont {Perfetto}},
  \bibinfo {author} {\bibfnamefont {G.}~\bibnamefont {Stefanucci}},\ and\
  \bibinfo {author} {\bibfnamefont {A.}~\bibnamefont {Marini}},\ }\href
  {https://doi.org/10.1140/epjb/e2018-90126-5} {\bibfield  {journal} {\bibinfo
  {journal} {The European Physical Journal B}\ }\textbf {\bibinfo {volume}
  {91}},\ \bibinfo {pages} {171} (\bibinfo {year} {2018})}\BibitemShut
  {NoStop}%
\bibitem [{\citenamefont {Marini}\ \emph {et~al.}(2022)\citenamefont {Marini},
  \citenamefont {Perfetto},\ and\ \citenamefont {Stefanucci}}]{Marini2022}%
  \BibitemOpen
  \bibfield  {author} {\bibinfo {author} {\bibfnamefont {A.}~\bibnamefont
  {Marini}}, \bibinfo {author} {\bibfnamefont {E.}~\bibnamefont {Perfetto}},\
  and\ \bibinfo {author} {\bibfnamefont {G.}~\bibnamefont {Stefanucci}},\
  }\href {https://doi.org/https://doi.org/10.1016/j.elspec.2022.147189}
  {\bibfield  {journal} {\bibinfo  {journal} {Journal of Electron Spectroscopy
  and Related Phenomena}\ }\textbf {\bibinfo {volume} {257}},\ \bibinfo {pages}
  {147189} (\bibinfo {year} {2022})}\BibitemShut {NoStop}%
\bibitem [{\citenamefont {Dauth}\ \emph {et~al.}(2014)\citenamefont {Dauth},
  \citenamefont {Wiessner}, \citenamefont {Feyer}, \citenamefont {Sch\"oll},
  \citenamefont {Puschnig}, \citenamefont {Reinert},\ and\ \citenamefont
  {K\"ummel}}]{Dauth2014}%
  \BibitemOpen
  \bibfield  {author} {\bibinfo {author} {\bibfnamefont {M.}~\bibnamefont
  {Dauth}}, \bibinfo {author} {\bibfnamefont {M.}~\bibnamefont {Wiessner}},
  \bibinfo {author} {\bibfnamefont {V.}~\bibnamefont {Feyer}}, \bibinfo
  {author} {\bibfnamefont {A.}~\bibnamefont {Sch\"oll}}, \bibinfo {author}
  {\bibfnamefont {P.}~\bibnamefont {Puschnig}}, \bibinfo {author}
  {\bibfnamefont {F.}~\bibnamefont {Reinert}},\ and\ \bibinfo {author}
  {\bibfnamefont {S.}~\bibnamefont {K\"ummel}},\ }\href
  {https://doi.org/10.1088/1367-2630/16/10/103005} {\bibfield  {journal}
  {\bibinfo  {journal} {New J. Phys.}\ }\textbf {\bibinfo {volume} {16}},\
  \bibinfo {pages} {103005} (\bibinfo {year} {2014})}\BibitemShut {NoStop}%
\bibitem [{\citenamefont {Damascelli}(2004)}]{Damascelli2004}%
  \BibitemOpen
  \bibfield  {author} {\bibinfo {author} {\bibfnamefont {A.}~\bibnamefont
  {Damascelli}},\ }\href@noop {} {\bibfield  {journal} {\bibinfo  {journal}
  {Phys. Scr.}\ }\textbf {\bibinfo {volume} {T109}},\ \bibinfo {pages} {61}
  (\bibinfo {year} {2004})}\BibitemShut {NoStop}%
\bibitem [{\citenamefont {Truhlar}\ \emph {et~al.}(2019)\citenamefont
  {Truhlar}, \citenamefont {Hiberty}, \citenamefont {Shaik}, \citenamefont
  {Gordon},\ and\ \citenamefont {Danovich}}]{Truhlar2019}%
  \BibitemOpen
  \bibfield  {author} {\bibinfo {author} {\bibfnamefont {D.~G.}\ \bibnamefont
  {Truhlar}}, \bibinfo {author} {\bibfnamefont {P.~C.}\ \bibnamefont
  {Hiberty}}, \bibinfo {author} {\bibfnamefont {S.}~\bibnamefont {Shaik}},
  \bibinfo {author} {\bibfnamefont {M.~S.}\ \bibnamefont {Gordon}},\ and\
  \bibinfo {author} {\bibfnamefont {D.}~\bibnamefont {Danovich}},\ }\href
  {https://doi.org/10.1002/ange.201904609} {\bibfield  {journal} {\bibinfo
  {journal} {Angewandte Chemie}\ }\textbf {\bibinfo {volume} {131}},\ \bibinfo
  {pages} {12460} (\bibinfo {year} {2019})}\BibitemShut {NoStop}%
\bibitem [{\citenamefont {Krylov}(2020)}]{Krylov2020}%
  \BibitemOpen
  \bibfield  {author} {\bibinfo {author} {\bibfnamefont {A.~I.}\ \bibnamefont
  {Krylov}},\ }\href {https://doi.org/10.1063/5.0018597} {\bibfield  {journal}
  {\bibinfo  {journal} {J. Chem. Phys.}\ }\textbf {\bibinfo {volume} {153}},\
  \bibinfo {pages} {080901} (\bibinfo {year} {2020})}\BibitemShut {NoStop}%
\bibitem [{\citenamefont {Galvani}\ \emph {et~al.}(2016)\citenamefont
  {Galvani}, \citenamefont {Paleari}, \citenamefont {Miranda}, \citenamefont
  {Molina-S\'anchez}, \citenamefont {Wirtz}, \citenamefont {Latil},
  \citenamefont {Amara},\ and\ \citenamefont {Ducastelle}}]{Galvani2016}%
  \BibitemOpen
  \bibfield  {author} {\bibinfo {author} {\bibfnamefont {T.}~\bibnamefont
  {Galvani}}, \bibinfo {author} {\bibfnamefont {F.}~\bibnamefont {Paleari}},
  \bibinfo {author} {\bibfnamefont {H.~P.~C.}\ \bibnamefont {Miranda}},
  \bibinfo {author} {\bibfnamefont {A.}~\bibnamefont {Molina-S\'anchez}},
  \bibinfo {author} {\bibfnamefont {L.}~\bibnamefont {Wirtz}}, \bibinfo
  {author} {\bibfnamefont {S.}~\bibnamefont {Latil}}, \bibinfo {author}
  {\bibfnamefont {H.}~\bibnamefont {Amara}},\ and\ \bibinfo {author}
  {\bibfnamefont {F.~m.~c.}\ \bibnamefont {Ducastelle}},\ }\href
  {https://doi.org/10.1103/PhysRevB.94.125303} {\bibfield  {journal} {\bibinfo
  {journal} {Phys. Rev. B}\ }\textbf {\bibinfo {volume} {94}},\ \bibinfo
  {pages} {125303} (\bibinfo {year} {2016})}\BibitemShut {NoStop}%
\bibitem [{\citenamefont {Ur{\'i}a-{\'A}lvarez}\ \emph
  {et~al.}(2024)\citenamefont {Ur{\'i}a-{\'A}lvarez}, \citenamefont
  {Esteve-Paredes}, \citenamefont {Garc{\'i}a-Bl{\'a}zquez},\ and\
  \citenamefont {Palacios}}]{UriaAlvarez2024}%
  \BibitemOpen
  \bibfield  {author} {\bibinfo {author} {\bibfnamefont {A.~J.}\ \bibnamefont
  {Ur{\'i}a-{\'A}lvarez}}, \bibinfo {author} {\bibfnamefont {J.~J.}\
  \bibnamefont {Esteve-Paredes}}, \bibinfo {author} {\bibfnamefont
  {M.}~\bibnamefont {Garc{\'i}a-Bl{\'a}zquez}},\ and\ \bibinfo {author}
  {\bibfnamefont {J.~J.}\ \bibnamefont {Palacios}},\ }\href
  {https://doi.org/https://doi.org/10.1016/j.cpc.2023.109001} {\bibfield
  {journal} {\bibinfo  {journal} {Computer Physics Communications}\ }\textbf
  {\bibinfo {volume} {295}},\ \bibinfo {pages} {109001} (\bibinfo {year}
  {2024})}\BibitemShut {NoStop}%
\bibitem [{\citenamefont {Giannozzi}\ \emph {et~al.}(2009)\citenamefont
  {Giannozzi}, \citenamefont {Baroni}, \citenamefont {Bonini}, \citenamefont
  {Calandra}, \citenamefont {Car}, \citenamefont {Cavazzoni}, \citenamefont
  {Ceresoli}, \citenamefont {Chiarotti}, \citenamefont {Cococcioni},
  \citenamefont {Dabo}, \citenamefont {Corso}, \citenamefont {de~Gironcoli},
  \citenamefont {Fabris}, \citenamefont {Fratesi}, \citenamefont {Gebauer},
  \citenamefont {Gerstmann}, \citenamefont {Gougoussis}, \citenamefont
  {Kokalj}, \citenamefont {Lazzeri}, \citenamefont {Martin-Samos},
  \citenamefont {Marzari}, \citenamefont {Mauri}, \citenamefont {Mazzarello},
  \citenamefont {Paolini}, \citenamefont {Pasquarello}, \citenamefont
  {Paulatto}, \citenamefont {Sbraccia}, \citenamefont {Scandolo}, \citenamefont
  {Sclauzero}, \citenamefont {Seitsonen}, \citenamefont {Smogunov},
  \citenamefont {Umari},\ and\ \citenamefont {Wentzcovitch}}]{Giannozzi2009}%
  \BibitemOpen
  \bibfield  {author} {\bibinfo {author} {\bibfnamefont {P.}~\bibnamefont
  {Giannozzi}}, \bibinfo {author} {\bibfnamefont {S.}~\bibnamefont {Baroni}},
  \bibinfo {author} {\bibfnamefont {N.}~\bibnamefont {Bonini}}, \bibinfo
  {author} {\bibfnamefont {M.}~\bibnamefont {Calandra}}, \bibinfo {author}
  {\bibfnamefont {R.}~\bibnamefont {Car}}, \bibinfo {author} {\bibfnamefont
  {C.}~\bibnamefont {Cavazzoni}}, \bibinfo {author} {\bibfnamefont
  {D.}~\bibnamefont {Ceresoli}}, \bibinfo {author} {\bibfnamefont {G.~L.}\
  \bibnamefont {Chiarotti}}, \bibinfo {author} {\bibfnamefont {M.}~\bibnamefont
  {Cococcioni}}, \bibinfo {author} {\bibfnamefont {I.}~\bibnamefont {Dabo}},
  \bibinfo {author} {\bibfnamefont {A.~D.}\ \bibnamefont {Corso}}, \bibinfo
  {author} {\bibfnamefont {S.}~\bibnamefont {de~Gironcoli}}, \bibinfo {author}
  {\bibfnamefont {S.}~\bibnamefont {Fabris}}, \bibinfo {author} {\bibfnamefont
  {G.}~\bibnamefont {Fratesi}}, \bibinfo {author} {\bibfnamefont
  {R.}~\bibnamefont {Gebauer}}, \bibinfo {author} {\bibfnamefont
  {U.}~\bibnamefont {Gerstmann}}, \bibinfo {author} {\bibfnamefont
  {C.}~\bibnamefont {Gougoussis}}, \bibinfo {author} {\bibfnamefont
  {A.}~\bibnamefont {Kokalj}}, \bibinfo {author} {\bibfnamefont
  {M.}~\bibnamefont {Lazzeri}}, \bibinfo {author} {\bibfnamefont
  {L.}~\bibnamefont {Martin-Samos}}, \bibinfo {author} {\bibfnamefont
  {N.}~\bibnamefont {Marzari}}, \bibinfo {author} {\bibfnamefont
  {F.}~\bibnamefont {Mauri}}, \bibinfo {author} {\bibfnamefont
  {R.}~\bibnamefont {Mazzarello}}, \bibinfo {author} {\bibfnamefont
  {S.}~\bibnamefont {Paolini}}, \bibinfo {author} {\bibfnamefont
  {A.}~\bibnamefont {Pasquarello}}, \bibinfo {author} {\bibfnamefont
  {L.}~\bibnamefont {Paulatto}}, \bibinfo {author} {\bibfnamefont
  {C.}~\bibnamefont {Sbraccia}}, \bibinfo {author} {\bibfnamefont
  {S.}~\bibnamefont {Scandolo}}, \bibinfo {author} {\bibfnamefont
  {G.}~\bibnamefont {Sclauzero}}, \bibinfo {author} {\bibfnamefont {A.~P.}\
  \bibnamefont {Seitsonen}}, \bibinfo {author} {\bibfnamefont {A.}~\bibnamefont
  {Smogunov}}, \bibinfo {author} {\bibfnamefont {P.}~\bibnamefont {Umari}},\
  and\ \bibinfo {author} {\bibfnamefont {R.~M.}\ \bibnamefont {Wentzcovitch}},\
  }\href {https://doi.org/10.1088/0953-8984/21/39/395502} {\bibfield  {journal}
  {\bibinfo  {journal} {J. Phys. Condens Matter}\ }\textbf {\bibinfo {volume}
  {21}},\ \bibinfo {pages} {395502} (\bibinfo {year} {2009})}\BibitemShut
  {NoStop}%
\bibitem [{\citenamefont {Giannozzi}\ \emph {et~al.}(2017)\citenamefont
  {Giannozzi}, \citenamefont {Andreussi}, \citenamefont {Brumme}, \citenamefont
  {Bunau}, \citenamefont {Nardelli}, \citenamefont {Calandra}, \citenamefont
  {Car}, \citenamefont {Cavazzoni}, \citenamefont {Ceresoli}, \citenamefont
  {Cococcioni}, \citenamefont {Colonna}, \citenamefont {Carnimeo},
  \citenamefont {Corso}, \citenamefont {de~Gironcoli}, \citenamefont {Delugas},
  \citenamefont {DiStasio}, \citenamefont {Ferretti}, \citenamefont {Floris},
  \citenamefont {Fratesi}, \citenamefont {Fugallo}, \citenamefont {Gebauer},
  \citenamefont {Gerstmann}, \citenamefont {Giustino}, \citenamefont {Gorni},
  \citenamefont {Jia}, \citenamefont {Kawamura}, \citenamefont {Ko},
  \citenamefont {Kokalj}, \citenamefont {Küçükbenli}, \citenamefont
  {Lazzeri}, \citenamefont {Marsili}, \citenamefont {Marzari}, \citenamefont
  {Mauri}, \citenamefont {Nguyen}, \citenamefont {Nguyen}, \citenamefont {de-la
  Roza}, \citenamefont {Paulatto}, \citenamefont {Poncé}, \citenamefont
  {Rocca}, \citenamefont {Sabatini}, \citenamefont {Santra}, \citenamefont
  {Schlipf}, \citenamefont {Seitsonen}, \citenamefont {Smogunov}, \citenamefont
  {Timrov}, \citenamefont {Thonhauser}, \citenamefont {Umari}, \citenamefont
  {Vast}, \citenamefont {Wu},\ and\ \citenamefont {Baroni}}]{Giannozzi2017}%
  \BibitemOpen
  \bibfield  {author} {\bibinfo {author} {\bibfnamefont {P.}~\bibnamefont
  {Giannozzi}}, \bibinfo {author} {\bibfnamefont {O.}~\bibnamefont
  {Andreussi}}, \bibinfo {author} {\bibfnamefont {T.}~\bibnamefont {Brumme}},
  \bibinfo {author} {\bibfnamefont {O.}~\bibnamefont {Bunau}}, \bibinfo
  {author} {\bibfnamefont {M.~B.}\ \bibnamefont {Nardelli}}, \bibinfo {author}
  {\bibfnamefont {M.}~\bibnamefont {Calandra}}, \bibinfo {author}
  {\bibfnamefont {R.}~\bibnamefont {Car}}, \bibinfo {author} {\bibfnamefont
  {C.}~\bibnamefont {Cavazzoni}}, \bibinfo {author} {\bibfnamefont
  {D.}~\bibnamefont {Ceresoli}}, \bibinfo {author} {\bibfnamefont
  {M.}~\bibnamefont {Cococcioni}}, \bibinfo {author} {\bibfnamefont
  {N.}~\bibnamefont {Colonna}}, \bibinfo {author} {\bibfnamefont
  {I.}~\bibnamefont {Carnimeo}}, \bibinfo {author} {\bibfnamefont {A.~D.}\
  \bibnamefont {Corso}}, \bibinfo {author} {\bibfnamefont {S.}~\bibnamefont
  {de~Gironcoli}}, \bibinfo {author} {\bibfnamefont {P.}~\bibnamefont
  {Delugas}}, \bibinfo {author} {\bibfnamefont {R.~A.}\ \bibnamefont
  {DiStasio}}, \bibinfo {author} {\bibfnamefont {A.}~\bibnamefont {Ferretti}},
  \bibinfo {author} {\bibfnamefont {A.}~\bibnamefont {Floris}}, \bibinfo
  {author} {\bibfnamefont {G.}~\bibnamefont {Fratesi}}, \bibinfo {author}
  {\bibfnamefont {G.}~\bibnamefont {Fugallo}}, \bibinfo {author} {\bibfnamefont
  {R.}~\bibnamefont {Gebauer}}, \bibinfo {author} {\bibfnamefont
  {U.}~\bibnamefont {Gerstmann}}, \bibinfo {author} {\bibfnamefont
  {F.}~\bibnamefont {Giustino}}, \bibinfo {author} {\bibfnamefont
  {T.}~\bibnamefont {Gorni}}, \bibinfo {author} {\bibfnamefont
  {J.}~\bibnamefont {Jia}}, \bibinfo {author} {\bibfnamefont {M.}~\bibnamefont
  {Kawamura}}, \bibinfo {author} {\bibfnamefont {H.-Y.}\ \bibnamefont {Ko}},
  \bibinfo {author} {\bibfnamefont {A.}~\bibnamefont {Kokalj}}, \bibinfo
  {author} {\bibfnamefont {E.}~\bibnamefont {Küçükbenli}}, \bibinfo {author}
  {\bibfnamefont {M.}~\bibnamefont {Lazzeri}}, \bibinfo {author} {\bibfnamefont
  {M.}~\bibnamefont {Marsili}}, \bibinfo {author} {\bibfnamefont
  {N.}~\bibnamefont {Marzari}}, \bibinfo {author} {\bibfnamefont
  {F.}~\bibnamefont {Mauri}}, \bibinfo {author} {\bibfnamefont {N.~L.}\
  \bibnamefont {Nguyen}}, \bibinfo {author} {\bibfnamefont {H.-V.}\
  \bibnamefont {Nguyen}}, \bibinfo {author} {\bibfnamefont {A.~O.}\
  \bibnamefont {de-la Roza}}, \bibinfo {author} {\bibfnamefont
  {L.}~\bibnamefont {Paulatto}}, \bibinfo {author} {\bibfnamefont
  {S.}~\bibnamefont {Poncé}}, \bibinfo {author} {\bibfnamefont
  {D.}~\bibnamefont {Rocca}}, \bibinfo {author} {\bibfnamefont
  {R.}~\bibnamefont {Sabatini}}, \bibinfo {author} {\bibfnamefont
  {B.}~\bibnamefont {Santra}}, \bibinfo {author} {\bibfnamefont
  {M.}~\bibnamefont {Schlipf}}, \bibinfo {author} {\bibfnamefont {A.~P.}\
  \bibnamefont {Seitsonen}}, \bibinfo {author} {\bibfnamefont {A.}~\bibnamefont
  {Smogunov}}, \bibinfo {author} {\bibfnamefont {I.}~\bibnamefont {Timrov}},
  \bibinfo {author} {\bibfnamefont {T.}~\bibnamefont {Thonhauser}}, \bibinfo
  {author} {\bibfnamefont {P.}~\bibnamefont {Umari}}, \bibinfo {author}
  {\bibfnamefont {N.}~\bibnamefont {Vast}}, \bibinfo {author} {\bibfnamefont
  {X.}~\bibnamefont {Wu}},\ and\ \bibinfo {author} {\bibfnamefont
  {S.}~\bibnamefont {Baroni}},\ }\href
  {https://doi.org/10.1088/1361-648X/aa8f79} {\bibfield  {journal} {\bibinfo
  {journal} {Journal of Physics: Condensed Matter}\ }\textbf {\bibinfo {volume}
  {29}},\ \bibinfo {pages} {465901} (\bibinfo {year} {2017})}\BibitemShut
  {NoStop}%
\bibitem [{\citenamefont {Giannozzi}\ \emph {et~al.}(2020)\citenamefont
  {Giannozzi}, \citenamefont {Baseggio}, \citenamefont {Bonfà}, \citenamefont
  {Brunato}, \citenamefont {Car}, \citenamefont {Carnimeo}, \citenamefont
  {Cavazzoni}, \citenamefont {de~Gironcoli}, \citenamefont {Delugas},
  \citenamefont {Ferrari~Ruffino}, \citenamefont {Ferretti}, \citenamefont
  {Marzari}, \citenamefont {Timrov}, \citenamefont {Urru},\ and\ \citenamefont
  {Baroni}}]{Giannozzi2020}%
  \BibitemOpen
  \bibfield  {author} {\bibinfo {author} {\bibfnamefont {P.}~\bibnamefont
  {Giannozzi}}, \bibinfo {author} {\bibfnamefont {O.}~\bibnamefont {Baseggio}},
  \bibinfo {author} {\bibfnamefont {P.}~\bibnamefont {Bonfà}}, \bibinfo
  {author} {\bibfnamefont {D.}~\bibnamefont {Brunato}}, \bibinfo {author}
  {\bibfnamefont {R.}~\bibnamefont {Car}}, \bibinfo {author} {\bibfnamefont
  {I.}~\bibnamefont {Carnimeo}}, \bibinfo {author} {\bibfnamefont
  {C.}~\bibnamefont {Cavazzoni}}, \bibinfo {author} {\bibfnamefont
  {S.}~\bibnamefont {de~Gironcoli}}, \bibinfo {author} {\bibfnamefont
  {P.}~\bibnamefont {Delugas}}, \bibinfo {author} {\bibfnamefont
  {F.}~\bibnamefont {Ferrari~Ruffino}}, \bibinfo {author} {\bibfnamefont
  {A.}~\bibnamefont {Ferretti}}, \bibinfo {author} {\bibfnamefont
  {N.}~\bibnamefont {Marzari}}, \bibinfo {author} {\bibfnamefont
  {I.}~\bibnamefont {Timrov}}, \bibinfo {author} {\bibfnamefont
  {A.}~\bibnamefont {Urru}},\ and\ \bibinfo {author} {\bibfnamefont
  {S.}~\bibnamefont {Baroni}},\ }\href {https://doi.org/10.1063/5.0005082}
  {\bibfield  {journal} {\bibinfo  {journal} {The Journal of Chemical Physics}\
  }\textbf {\bibinfo {volume} {152}},\ \bibinfo {pages} {154105} (\bibinfo
  {year} {2020})},\ \Eprint
  {https://arxiv.org/abs/https://pubs.aip.org/aip/jcp/article-pdf/doi/10.1063/5.0005082/16721881/154105\_1\_online.pdf}
  {https://pubs.aip.org/aip/jcp/article-pdf/doi/10.1063/5.0005082/16721881/154105\_1\_online.pdf}
  \BibitemShut {NoStop}%
\bibitem [{\citenamefont {Perdew}\ \emph {et~al.}(1996)\citenamefont {Perdew},
  \citenamefont {Burke},\ and\ \citenamefont {Ernzerhof}}]{Perdew1996}%
  \BibitemOpen
  \bibfield  {author} {\bibinfo {author} {\bibfnamefont {J.~P.}\ \bibnamefont
  {Perdew}}, \bibinfo {author} {\bibfnamefont {K.}~\bibnamefont {Burke}},\ and\
  \bibinfo {author} {\bibfnamefont {M.}~\bibnamefont {Ernzerhof}},\ }\href
  {https://doi.org/10.1103/PhysRevLett.77.3865} {\bibfield  {journal} {\bibinfo
   {journal} {Phys. Rev. Lett.}\ }\textbf {\bibinfo {volume} {77}},\ \bibinfo
  {pages} {3865} (\bibinfo {year} {1996})}\BibitemShut {NoStop}%
\bibitem [{\citenamefont {Deslippe}\ \emph {et~al.}(2012)\citenamefont
  {Deslippe}, \citenamefont {Samsonidze}, \citenamefont {Strubbe},
  \citenamefont {Jain}, \citenamefont {Cohen},\ and\ \citenamefont
  {Louie}}]{Deslippe2012}%
  \BibitemOpen
  \bibfield  {author} {\bibinfo {author} {\bibfnamefont {J.}~\bibnamefont
  {Deslippe}}, \bibinfo {author} {\bibfnamefont {G.}~\bibnamefont
  {Samsonidze}}, \bibinfo {author} {\bibfnamefont {D.~A.}\ \bibnamefont
  {Strubbe}}, \bibinfo {author} {\bibfnamefont {M.}~\bibnamefont {Jain}},
  \bibinfo {author} {\bibfnamefont {M.~L.}\ \bibnamefont {Cohen}},\ and\
  \bibinfo {author} {\bibfnamefont {S.~G.}\ \bibnamefont {Louie}},\ }\href
  {https://doi.org/10.1016/j.cpc.2011.12.006} {\bibfield  {journal} {\bibinfo
  {journal} {Computer Physics Communications}\ }\textbf {\bibinfo {volume}
  {183}},\ \bibinfo {pages} {1269} (\bibinfo {year} {2012})}\BibitemShut
  {NoStop}%
\bibitem [{\citenamefont {Kirchhoff}\ \emph {et~al.}(2022)\citenamefont
  {Kirchhoff}, \citenamefont {Deilmann}, \citenamefont {Kr\"uger},\ and\
  \citenamefont {Rohlfing}}]{Kirchhoff2022}%
  \BibitemOpen
  \bibfield  {author} {\bibinfo {author} {\bibfnamefont {A.}~\bibnamefont
  {Kirchhoff}}, \bibinfo {author} {\bibfnamefont {T.}~\bibnamefont {Deilmann}},
  \bibinfo {author} {\bibfnamefont {P.}~\bibnamefont {Kr\"uger}},\ and\
  \bibinfo {author} {\bibfnamefont {M.}~\bibnamefont {Rohlfing}},\ }\href
  {https://doi.org/10.1103/PhysRevB.106.045118} {\bibfield  {journal} {\bibinfo
   {journal} {Phys. Rev. B}\ }\textbf {\bibinfo {volume} {106}},\ \bibinfo
  {pages} {045118} (\bibinfo {year} {2022})}\BibitemShut {NoStop}%
\bibitem [{\citenamefont {Wannier}(1937)}]{wannier1937structure}%
  \BibitemOpen
  \bibfield  {author} {\bibinfo {author} {\bibfnamefont {G.~H.}\ \bibnamefont
  {Wannier}},\ }\href@noop {} {\bibfield  {journal} {\bibinfo  {journal}
  {Physical Review}\ }\textbf {\bibinfo {volume} {52}},\ \bibinfo {pages} {191}
  (\bibinfo {year} {1937})}\BibitemShut {NoStop}%
\bibitem [{\citenamefont {Weinelt}\ \emph {et~al.}(2004)\citenamefont
  {Weinelt}, \citenamefont {Kutschera}, \citenamefont {Fauster},\ and\
  \citenamefont {Rohlfing}}]{Weinelt2004}%
  \BibitemOpen
  \bibfield  {author} {\bibinfo {author} {\bibfnamefont {M.}~\bibnamefont
  {Weinelt}}, \bibinfo {author} {\bibfnamefont {M.}~\bibnamefont {Kutschera}},
  \bibinfo {author} {\bibfnamefont {T.}~\bibnamefont {Fauster}},\ and\ \bibinfo
  {author} {\bibfnamefont {M.}~\bibnamefont {Rohlfing}},\ }\href
  {https://doi.org/10.1103/PhysRevLett.92.126801} {\bibfield  {journal}
  {\bibinfo  {journal} {Phys. Rev. Lett.}\ }\textbf {\bibinfo {volume} {92}},\
  \bibinfo {pages} {126801} (\bibinfo {year} {2004})}\BibitemShut {NoStop}%
\bibitem [{\citenamefont {Karni}\ \emph {et~al.}(2022)\citenamefont {Karni},
  \citenamefont {Barré}, \citenamefont {Pareek}, \citenamefont {Georgaras},
  \citenamefont {Man}, \citenamefont {Sahoo}, \citenamefont {Bacon},
  \citenamefont {Zhu}, \citenamefont {Ribeiro}, \citenamefont {O’Beirne},
  \citenamefont {Hu}, \citenamefont {Al-Mahboob}, \citenamefont {Abdelrasoul},
  \citenamefont {Chan}, \citenamefont {Karmakar}, \citenamefont {Winchester},
  \citenamefont {Kim}, \citenamefont {Watanabe}, \citenamefont {Taniguchi},
  \citenamefont {Barmak}, \citenamefont {Madéo}, \citenamefont {da~Jornada},
  \citenamefont {Heinz},\ and\ \citenamefont {Dani}}]{Karni2022}%
  \BibitemOpen
  \bibfield  {author} {\bibinfo {author} {\bibfnamefont {O.}~\bibnamefont
  {Karni}}, \bibinfo {author} {\bibfnamefont {E.}~\bibnamefont {Barré}},
  \bibinfo {author} {\bibfnamefont {V.}~\bibnamefont {Pareek}}, \bibinfo
  {author} {\bibfnamefont {J.~D.}\ \bibnamefont {Georgaras}}, \bibinfo {author}
  {\bibfnamefont {M.~K.~L.}\ \bibnamefont {Man}}, \bibinfo {author}
  {\bibfnamefont {C.}~\bibnamefont {Sahoo}}, \bibinfo {author} {\bibfnamefont
  {D.~R.}\ \bibnamefont {Bacon}}, \bibinfo {author} {\bibfnamefont
  {X.}~\bibnamefont {Zhu}}, \bibinfo {author} {\bibfnamefont {H.~B.}\
  \bibnamefont {Ribeiro}}, \bibinfo {author} {\bibfnamefont {A.~L.}\
  \bibnamefont {O’Beirne}}, \bibinfo {author} {\bibfnamefont
  {J.}~\bibnamefont {Hu}}, \bibinfo {author} {\bibfnamefont {A.}~\bibnamefont
  {Al-Mahboob}}, \bibinfo {author} {\bibfnamefont {M.~M.~M.}\ \bibnamefont
  {Abdelrasoul}}, \bibinfo {author} {\bibfnamefont {N.~S.}\ \bibnamefont
  {Chan}}, \bibinfo {author} {\bibfnamefont {A.}~\bibnamefont {Karmakar}},
  \bibinfo {author} {\bibfnamefont {A.~J.}\ \bibnamefont {Winchester}},
  \bibinfo {author} {\bibfnamefont {B.}~\bibnamefont {Kim}}, \bibinfo {author}
  {\bibfnamefont {K.}~\bibnamefont {Watanabe}}, \bibinfo {author}
  {\bibfnamefont {T.}~\bibnamefont {Taniguchi}}, \bibinfo {author}
  {\bibfnamefont {K.}~\bibnamefont {Barmak}}, \bibinfo {author} {\bibfnamefont
  {J.}~\bibnamefont {Madéo}}, \bibinfo {author} {\bibfnamefont {F.~H.}\
  \bibnamefont {da~Jornada}}, \bibinfo {author} {\bibfnamefont {T.~F.}\
  \bibnamefont {Heinz}},\ and\ \bibinfo {author} {\bibfnamefont {K.~M.}\
  \bibnamefont {Dani}},\ }\href {https://doi.org/10.1038/s41586-021-04360-y}
  {\bibfield  {journal} {\bibinfo  {journal} {Nature}\ }\textbf {\bibinfo
  {volume} {603}},\ \bibinfo {pages} {247} (\bibinfo {year}
  {2022})}\BibitemShut {NoStop}%
\bibitem [{\citenamefont {L\"uftner}\ \emph {et~al.}(2017)\citenamefont
  {L\"uftner}, \citenamefont {Wei{\ss}}, \citenamefont {Yang}, \citenamefont
  {Hurdax}, \citenamefont {Feyer}, \citenamefont {Gottwald}, \citenamefont
  {Koller}, \citenamefont {Soubatch}, \citenamefont {Puschnig}, \citenamefont
  {Ramsey},\ and\ \citenamefont {Tautz}}]{Lueftner2017}%
  \BibitemOpen
  \bibfield  {author} {\bibinfo {author} {\bibfnamefont {D.}~\bibnamefont
  {L\"uftner}}, \bibinfo {author} {\bibfnamefont {S.}~\bibnamefont {Wei{\ss}}},
  \bibinfo {author} {\bibfnamefont {X.}~\bibnamefont {Yang}}, \bibinfo {author}
  {\bibfnamefont {P.}~\bibnamefont {Hurdax}}, \bibinfo {author} {\bibfnamefont
  {V.}~\bibnamefont {Feyer}}, \bibinfo {author} {\bibfnamefont
  {A.}~\bibnamefont {Gottwald}}, \bibinfo {author} {\bibfnamefont
  {G.}~\bibnamefont {Koller}}, \bibinfo {author} {\bibfnamefont
  {S.}~\bibnamefont {Soubatch}}, \bibinfo {author} {\bibfnamefont
  {P.}~\bibnamefont {Puschnig}}, \bibinfo {author} {\bibfnamefont {M.~G.}\
  \bibnamefont {Ramsey}},\ and\ \bibinfo {author} {\bibfnamefont {F.~S.~S.}\
  \bibnamefont {Tautz}},\ }\href {https://doi.org/10.1103/PhysRevB.96.125402}
  {\bibfield  {journal} {\bibinfo  {journal} {Phys. Rev. B}\ }\textbf {\bibinfo
  {volume} {96}},\ \bibinfo {pages} {125402} (\bibinfo {year}
  {2017})}\BibitemShut {NoStop}%
\bibitem [{\citenamefont {Gierz}\ \emph {et~al.}(2011)\citenamefont {Gierz},
  \citenamefont {Henk}, \citenamefont {H\"ochst}, \citenamefont {Ast},\ and\
  \citenamefont {Kern}}]{Gierz2011}%
  \BibitemOpen
  \bibfield  {author} {\bibinfo {author} {\bibfnamefont {I.}~\bibnamefont
  {Gierz}}, \bibinfo {author} {\bibfnamefont {J.}~\bibnamefont {Henk}},
  \bibinfo {author} {\bibfnamefont {H.}~\bibnamefont {H\"ochst}}, \bibinfo
  {author} {\bibfnamefont {C.~R.}\ \bibnamefont {Ast}},\ and\ \bibinfo {author}
  {\bibfnamefont {K.}~\bibnamefont {Kern}},\ }\href
  {https://doi.org/10.1103/PhysRevB.83.121408} {\bibfield  {journal} {\bibinfo
  {journal} {Phys. Rev. B}\ }\textbf {\bibinfo {volume} {83}},\ \bibinfo
  {pages} {121408} (\bibinfo {year} {2011})}\BibitemShut {NoStop}%
\bibitem [{\citenamefont {Krasovskii}(2021)}]{Krasovskii2021}%
  \BibitemOpen
  \bibfield  {author} {\bibinfo {author} {\bibfnamefont {E.}~\bibnamefont
  {Krasovskii}},\ }\bibfield  {journal} {\bibinfo  {journal} {Nanomaterials}\
  }\textbf {\bibinfo {volume} {11}},\ \href
  {https://doi.org/10.3390/nano11051212} {10.3390/nano11051212} (\bibinfo
  {year} {2021})\BibitemShut {NoStop}%
\bibitem [{\citenamefont {Bosak}\ \emph {et~al.}(2006)\citenamefont {Bosak},
  \citenamefont {Serrano}, \citenamefont {Krisch}, \citenamefont {Watanabe},
  \citenamefont {Taniguchi},\ and\ \citenamefont {Kanda}}]{Bosak2006}%
  \BibitemOpen
  \bibfield  {author} {\bibinfo {author} {\bibfnamefont {A.}~\bibnamefont
  {Bosak}}, \bibinfo {author} {\bibfnamefont {J.}~\bibnamefont {Serrano}},
  \bibinfo {author} {\bibfnamefont {M.}~\bibnamefont {Krisch}}, \bibinfo
  {author} {\bibfnamefont {K.}~\bibnamefont {Watanabe}}, \bibinfo {author}
  {\bibfnamefont {T.}~\bibnamefont {Taniguchi}},\ and\ \bibinfo {author}
  {\bibfnamefont {H.}~\bibnamefont {Kanda}},\ }\href
  {https://doi.org/10.1103/PhysRevB.73.041402} {\bibfield  {journal} {\bibinfo
  {journal} {Phys. Rev. B}\ }\textbf {\bibinfo {volume} {73}},\ \bibinfo
  {pages} {041402} (\bibinfo {year} {2006})}\BibitemShut {NoStop}%
\bibitem [{\citenamefont {Sohier}\ \emph {et~al.}(2017)\citenamefont {Sohier},
  \citenamefont {Calandra},\ and\ \citenamefont {Mauri}}]{Sohier2017}%
  \BibitemOpen
  \bibfield  {author} {\bibinfo {author} {\bibfnamefont {T.}~\bibnamefont
  {Sohier}}, \bibinfo {author} {\bibfnamefont {M.}~\bibnamefont {Calandra}},\
  and\ \bibinfo {author} {\bibfnamefont {F.}~\bibnamefont {Mauri}},\ }\href
  {https://doi.org/10.1103/PhysRevB.96.075448} {\bibfield  {journal} {\bibinfo
  {journal} {Phys. Rev. B}\ }\textbf {\bibinfo {volume} {96}},\ \bibinfo
  {pages} {075448} (\bibinfo {year} {2017})}\BibitemShut {NoStop}%
\bibitem [{\citenamefont {da~Jornada}\ \emph {et~al.}(2017)\citenamefont
  {da~Jornada}, \citenamefont {Qiu},\ and\ \citenamefont
  {Louie}}]{da_Jornada2017}%
  \BibitemOpen
  \bibfield  {author} {\bibinfo {author} {\bibfnamefont {F.~H.}\ \bibnamefont
  {da~Jornada}}, \bibinfo {author} {\bibfnamefont {D.~Y.}\ \bibnamefont
  {Qiu}},\ and\ \bibinfo {author} {\bibfnamefont {S.~G.}\ \bibnamefont
  {Louie}},\ }\href {https://doi.org/10.1103/physrevb.95.035109} {\bibfield
  {journal} {\bibinfo  {journal} {Physical Review B}\ }\textbf {\bibinfo
  {volume} {95}},\ \bibinfo {pages} {035109} (\bibinfo {year}
  {2017})}\BibitemShut {NoStop}%
\end{thebibliography}%
\end{document}